\documentclass[11pt, letterpaper]{article}
\pdfoutput=1
\usepackage[authoryear,round]{natbib}
\usepackage{amsmath,amssymb,amsfonts,amsthm,bbm}
\usepackage{epic,eepic,epsfig,longtable}
\usepackage{multirow,verbatim}
\usepackage{array}
\usepackage{hyperref}

\usepackage{epsfig}
\usepackage{setspace}
\usepackage{CJK}
\usepackage{color}

\providecommand{\abs}[1]{\left\lvert#1\right\rvert}
\providecommand{\norm}[1]{\left\lVert#1\right\rVert}

\makeatletter
\def\singlespace{\def\baselinestretch{1}\@normalsize}

\makeatletter
\def\singlespace{\def\baselinestretch{1}\@normalsize}

\numberwithin{equation}{section}

\renewcommand{\hat}{\widehat}

\renewcommand{\hat}{\widehat}

\newcommand{\bfm}[1]{\ensuremath{\mathbf{#1}}}

   \def\bA{\bfm A}  
   \def\bB{\bfm B}  
   \def\bC{\bfm C}  
   \def\bD{\bfm D}  
\def\be{\bfm e}

\def\bh{\bfm h}   \def\bH{\bfm H}  
   \def\bI{\bfm I}

   \def\bL{\bfm L}  
   \def\bM{\bfm M}

     \def\PP{\mathbb{P}}
     
   \def\bR{\bfm R}  \def\RR{\mathbb{R}}
   \def\bS{\bfm S}  
   \def\bT{\bfm T}  
\def\bu{\bfm u}   \def\bU{\bfm U}  
\def\bv{\bfm v}   \def\bV{\bfm V}  
   \def\bW{\bfm W}  
\def\bx{\bfm x}   \def\bX{\bfm X}  
\def\by{\bfm y}   \def\bY{\bfm Y}  
\def\bz{\bfm z}   \def\bZ{\bfm Z}

 \def\cB{{\cal  B}}
 \def\cC{{\cal  C}}
 
 \def\cE{{\cal  E}}

 \def\cL{{\cal  L}}
 \def\cM{{\cal  M}}
 \def\cN{{\cal  N}}

 \def\cS{{\cal  S}}

 \def\cW{{\cal  W}}
 
 \def\cY{{\cal  Y}}
 
\def\bzero{\bfm 0}

\newcommand{\bfsym}[1]{\ensuremath{\boldsymbol{#1}}}

              \def\bGamma{\bfsym \Gamma}
            \def\bDelta {\bfsym {\Delta}}

 \def\btheta{\bfsym {\theta}}           \def\bTheta {\bfsym {\Theta}}
           \def\bepsilon{\bfsym \varepsilon}
              \def\bSigma{\bfsym \Sigma}

 \def\brho   {\bfsym {\rho}}
 
 \def\bxi{\bfsym {\xi}}
 
\DeclareMathOperator{\argmin}{argmin}

\DeclareMathOperator{\diag}{diag}
\DeclareMathOperator{\E}{E}

\DeclareMathOperator{\rank}{rank}

\DeclareMathOperator{\sgn}{sgn}

\DeclareMathOperator{\tr}{tr}

\def\today{\ifcase\month\or
  January\or February\or March\or April\or May\or June\or
  July\or August\or September\or October\or November\or December\fi
  \space\number\day, \number\year}

\newdimen\biblioindent    \biblioindent=30pt

 at 8truept

\def\sgn{\mbox{sgn}}

\newcommand{\beq}{\begin{equation}}
  \newcommand{\eeq}{\end{equation}}
\newcommand{\beqn}{\begin{eqnarray}}
  \newcommand{\eeqn}{\end{eqnarray}}
\newcommand{\beqnn}{\begin{eqnarray*}}
  \newcommand{\eeqnn}{\end{eqnarray*}}

\allowdisplaybreaks
\usepackage{verbatim}
\renewcommand{\baselinestretch}{1.66}
\newtheorem{lem}{Lemma}
\newtheorem{dfn}{Definition}
\newtheorem{thm}{Theorem}

\newtheorem{rmk}{Remark}
\newcounter{CondCounter}

\def \bbP      {\mathbb{P}}

\usepackage{amsmath}
\usepackage{amssymb}
\usepackage{graphicx}
\usepackage{tabularx}
\usepackage{textcomp}
\usepackage{float}
\usepackage{pgf,tikz}
\usepackage{subfig}
\usepackage{array}
\usepackage{dsfont}
\usepackage{verbatim}
\makeatletter

\newcommand{\Rmnum}[1]{\expandafter\@slowromancap\romannumeral #1@}
\makeatother

\newcommand\given[1][]{\:#1\vert\:}
\newcommand{\inn}[1]{\langle #1 \rangle}
\allowdisplaybreaks[1]
\def \vec	{\text{vec}}

\newcommand{\ltwonorm}[1]{\lVert#1\rVert_2}

\newcommand{\opnorm}[1]{\lVert#1\rVert_{op}}
\newcommand{\fnorm}[1]{\lVert#1\rVert_F}
\newcommand{\nnorm}[1]{\lVert#1\rVert_N}
\newcommand{\supnorm}[1]{ \lVert#1  \rVert_{\infty}}
\newcommand\independent{\protect\mathpalette{\protect\independenT}{\perp}}
\def\independenT#1#2{\mathrel{\rlap{$#1#2$}\mkern2mu{#1#2}}}

\begin{document}

\title{\bf Generalized High-Dimensional Trace Regression via Nuclear Norm Regularization\footnote{This paper is supported by NSF grants DMS-1406266, DMS-1662139, and DMS-1712591}}
\author{Jianqing Fan, Wenyan Gong, Ziwei Zhu}

\date{\today}
\maketitle

\abstract
We study the generalized trace regression with a near low-rank regression coefficient matrix, which extends notion of sparsity for regression coefficient vectors. Specifically, given a matrix covariate $\bX$, the probability density function $f(Y\vert \bX)=c(Y)\exp{(\phi^{-1}\left[-Y\eta^* + b(\eta^*)\right])}$, where $\eta^*=\tr({\bTheta^*}^T\bX)$. This model accommodates various types of responses and embraces many important problem setups such as reduced-rank regression, matrix regression that accommodates a panel of regressors, matrix completion, among others. We estimate $\bTheta^*$ through minimizing empirical negative log-likelihood plus nuclear norm penalty.   We first establish a general theory and then for each specific problem, we derive explicitly the statistical rate of the proposed estimator. They all match the minimax rates in the linear trace regression up to logarithmic factors. Numerical studies confirm the rates we established and demonstrate the advantage of generalized trace regression over linear trace regression when the response is dichotomous. We also show the benefit of incorporating nuclear norm regularization in dynamic stock return prediction and in image classification.

\section{Introduction}



In modern data analytics, the parameters of interest often exhibit high ambient dimensions but low intrinsic dimensions that can be exploited to circumvent the curse of dimensionality. One of the most illustrating examples is the sparse signal recovery through incorporating sparsity regularization into empirical risk minimization (\cite{tibshirani1996regression, chen2001atomic, fan2001variable}). As shown in the profound works (\cite{candes2007dantzig, fan2008sure, fan2011nonconcave, zou2008one, zhang2010nearly}, among others), the statistical rate of the appropriately regularized M-estimator has mere logarithmic dependence on the ambient dimension $d$. This implies that consistent signal recovery is feasible even when $d$ grows exponentially with respect to the sample size $n$. In econometrics, sparse models and methods have also been intensively studied and are proven to be powerful. For example, \cite{BCC12} studied estimation of optimal instruments under sparse high-dimensional models and showed that the instrumental variable (IV) estimator based on Lasso and post-Lasso methods enjoys root-n consistency and asymptotic normality. \cite{hansen2014instrumental} and
\cite{caner2015hybrid} investigated instrument selection using high-dimensional regularization methods. \cite{kock2015oracle} established oracle inequalities for high dimensional vector autoregressions and
\cite{CYZ15} applied group Lasso in threshold autoregressive models and established near-optimal rates in the estimation of threshold parameters.
\cite{belloni2017program} employed high-dimensional techniques for program evaluation and causal inference.

When the parameter of interest arises in the matrix form, elementwise sparsity is not the sole way of constraining model complexity; another structure that is exclusive to matrices comes into play: the rank. Low-rank matrices have much fewer degrees of freedom than its ambient dimensions $d_1\cdot d_2$. To determine a rank-$r$ matrix $\bTheta \in \RR^{d_1 \times d_2}$, we only need $r$ left and right singular vectors and $r$ singular values, which correspond to $r(d_1 + d_2 - 1)$ degrees of freedom, without accounting the orthogonality. As a novel regularization approach, low-rankness motivates matrix representations of the parameters of interest in various statistical and econometric models. If we rearrange the coefficient in the traditional linear model as a matrix, we obtain the so-called trace regression model:
\begin{equation}
\label{eq:1.1}
	Y=\tr (\bTheta^{*T}\bX)+\epsilon,
\end{equation}
where $\tr(\cdot)$ denotes the trace, $\bX$ is a matrix of explanatory variables, $\bTheta^*\in\mathbb{R}^{d_1\times d_2}$ is the matrix of regression coefficients, $Y$ is the response and $\epsilon$ is the noise. 
In predictive econometric applications, $\bX$ can be a large panel of time series data such as stock returns or macroeconomic variables \citep{stock2002forecasting, ludvigson2009macro}, whereas in statistical machine learning $\bX$ can be images.
The rank of a matrix is controlled by the $\ell_q$-norm for $q\in[0,1)$ of its singular values:
\begin{equation}\label{eq:222}
    \cB_q(\bTheta^*) := \sum\limits_{j=1}^{d_1\wedge d_2}\sigma_j(\bTheta^*)^q\leq \rho,
\end{equation}
where $\sigma_j(\bTheta^*)$ is the $j$th largest singular value of $\bTheta^*$, and $\rho$ is a positive constant that can grow to infinity. Note that when $q=0$, it controls the rank of $\bTheta^*$ at $\rho$. Trace regression is a natural model  for matrix-type covariates, such as the panel data, images, genomics microarrays, etc. In addition, particular forms of $\bX$ can reduce trace regression to several well-known problem setups. For example, when $\bX$ contains only a column and the response $Y$ is multivariate, \eqref{eq:1.1} becomes reduced-rank regression model (\cite{And51}, \cite{Ize75}). When $\bX\in\mathbb{R}^{d_1\times d_2}$ is a singleton in the sense that all entries of $\bX$ are zeros except for one entry that equals one, \eqref{eq:1.1} characterizes the matrix completion problem in item response problems and online recommendation systems. We will specify these problems later.



To explore the low rank structure of $\bTheta^*$ in (\ref{eq:1.1}), a natural approach is the penalized least-squares with the nuclear norm penalty. Specifically, consider the following optimization problem.
\begin{equation}\label{eq:estimator}
\hat{\bTheta}=\argmin\left\{\frac{1}{n}\sum_{i=1}^n \left(\inn{\bTheta,\bX_i}-Y_i\right)^2+\lambda\norm{\bTheta}_N\right\},
\end{equation}
where  $\|\bTheta\|_N = \sum\limits_{j=1}^{d_1\wedge d_2} \sigma_j(\bTheta)$ is the nuclear norm of $\bTheta$. As $\ell_1$-norm regularization yields sparse estimators, nuclear norm regularization enforces the solution to have sparse singular values, in other words, to be low-rank. Recent literatures have rigorously studied the statistical properties of $\widehat\bTheta$. \cite{negahban2011estimation} and \cite{koltchinskii2011nuclear} derived the statistical error rate of $\widehat \bTheta $ when $\epsilon$ is sub-Gaussian.  \cite{fan2016shrinkage} introduced a shrinkage principle to handle heavy-tailed noise and achieved the same statistical error rate as \cite{negahban2011estimation} when the noise has merely bounded second moments.

However, \eqref{eq:1.1} does not accomodate categorical responses, which is ubiquitous in pragmatic settings. For example, in P2P microfinance, platforms like Kiva seek potential pairs of lenders and borrowers to create loans. The analysis is based on a large binary matrix with the rows correspondent to the lenders and columns correspondent to the borrowers. Entry $(i,j)$ of the matrix is either checked, meaning that lender $i$ endorses an loan to borrower $j$, or missing, meaning that lender $i$ is not interested in borrower $j$ or has not seen the request of borrower $j$. The specific amount of the loan is inaccessible due to privacy concern, thus leading to the binary nature of the response (\cite{LLC14}). Another example is the famous Netflix Challenge. There, people are given a large rating matrix with the rows representing the customers and the columns representing the movies. Most of its entries are missing and the aim is to infer these missing ratings based on the observed ones. Since the Netflix adopts a five-star movie rating system, the response is categorical with only five levels. This kind of matrix completion problems for item response arise also frequently in other economic surveys, similar to the aforementioned P2P microfinance.  
These problem setups with categorical responses motivate us to consider the generalized trace regression model.

Suppose that the response $Y$ follows a distribution from the following exponential family:
\begin{equation}\label{eq:model}
f_n(\bY;X,\beta^*)=\prod\limits_{i=1}^n f(Y_i;\eta_i^*)=\prod\limits_{i=1}^n \left\{c(Y_i)\exp{\left(\frac{Y_i\eta_i^*-b(\eta_i^*)}{\phi}\right)}\right\},
\end{equation}
where $\eta_i^*=\text{tr}(\bTheta^{*T}\bX_i)=\inn{\bTheta^*,\bX_i}$ is the linear predictor, $\phi$ is a constant and $c(\cdot)$ and $b(\cdot)$ are known functions. The negative log-likelihood corresponding to (\ref{eq:model}) is given, up to an affine transformation, by
\begin{equation}\label{eq:negloglike}
\cL_n(\bTheta)
=\frac{1}{n}\sum\limits_{i=1}^n\left[-Y_i\inn{\bTheta,\bX_i}+b(\inn{\bTheta,\bX_i})\right]
\end{equation}
and the gradient and Hessian of $\cL_n(\bTheta)$ are respectively
\begin{equation}\label{eq:gradhessian}
\begin{split}
&\nabla\cL_n(\bTheta)=\frac{1}{n}\sum\limits_{i=1}^n \left[b'(\eta_i)-Y_i\right]\bX_i=\frac{1}{n}\sum\limits_{i=1}^n \left[b'(\inn{\bTheta,\bX_i})-Y_i\right]\bX_i \\
& \widehat \bH(\bTheta) := \nabla^2\cL_n(\bTheta)=\frac{1}{n}\sum\limits_{i=1}^n b''(\inn{\bTheta,\bX_i})\vec(\bX_i)\vec(\bX_i)^T.
\end{split}
\end{equation}
To estimate $\bTheta^*$, we recruit the following M-estimator that minimizes the negative log-likelihood plus nuclear norm penalty.
\begin{equation}
\label{eq:1.7}
\hat{\bTheta}=\argmin_{\bTheta \in \RR^{d_1 \times d_2}} \left\{\frac{1}{n}\sum_{i=1}^n \left[b(\inn{\bTheta,\bX_i})-Y_i\inn{\bTheta,\bX_i}\right]+\lambda\norm{\bTheta}_N\right\}.
\end{equation}
This is a high-dimensional convex optimization problem.  We will discuss the algorithms for computing \eqref{eq:1.7} in the simulation section.

Related to our work is the matrix completion problem with binary entry, i.e., 1-bit matrix completion, which is a specific example of our generalized trace regression and has direct application in predicting aforementioned P2P microfinance. Therein entry $(i, j)$ of the matrix is modeled as a response from a logistic regression or probit regression with parameter $\bTheta^*_{ij}$ and information of each responded items is related through the low-rank assumption of $\bTheta^*$.
Previous works studied the estimation of $\bTheta^*$ by minimizing the negative log-likelihood function under the constraint of max-norm (\cite{cai2013max}), nuclear norm (\cite{davenport20141}) and rank (\cite{bhaskar20151}). 
There are also some works in 1-bit compressed sensing to recover sparse signal vectors \citep{gupta2010sample, plan2013one, plan2013robust}.  Nevertheless, we did not find any work in the generality that we are dealing with, which fits matrix-type explanatory variables and various types of dependent variables.

In this paper, we establish a unified framework for statistical analysis of $\widehat\bTheta$ in \eqref{eq:1.7} under the generalized trace regression model. As showcases of the applications of our general theory, we focus on three problem setups: generalized matrix regression, reduced-rank regression and one-bit matrix completion. We explicitly derive statistical rate of $\widehat \bTheta$ under these three problem setups respectively. It is worth noting that for one-bit matrix completion, our statistical rate is sharper than that in \cite{davenport20141}.
We also conduct numerical experiments on both simulated and real data to verify the established rate and illustrate the advantage of using the generalized trace regression over the vanilla trace regression when categorical responses occur.



The paper is organized as follows. In Section \ref{sec:2}, we specify the problem setups and present the statistical rates of $\widehat\bTheta$ under generalized matrix regression, reduced-rank regression and one-bit matrix completion respectively. In Section \ref{sec:3}, we present simulation results to back up our theoretical results from Section \ref{sec:2} and to demonstrate superiority of generalized trace regression over the standard one. In Section \ref{sec:4}, we use real data to display the improvement brought by nuclear norm regularization in return prediction and image classification.

\section{Main Results}\label{sec:2}

\subsection{Notation}


We use regular letters for random variables, bold lower case letters for random vectors and bold upper case letter for matrices. For a function $f(\cdot)$, we use $f'(\cdot)$, $f''(\cdot)$ and $f'''(\cdot)$ to denote its first, second and third order derivative.
For sequences $\{a_i\}_{i=1}^{\infty}$ and $\{b_i\}_{i=1}^{\infty}$, we say $a_i=O(b_i)$ if there exists a constant $c>0$ such that $a_i/b_i<c$ for $1\leq i<\infty$, and we say $a_i=\Omega(b_i)$ if there exists a constant $c>0$ such that $a_i/b_i \ge c$ for $1\leq i<\infty$.
For a random variable $x$, we denote its sub-Gaussian norm as $\norm{x}_{\Psi_2} := \sup_{p\geq 1}\left(\mathbb{E}\abs{x}^p\right)^{1/p}/\sqrt{p}$ and its sub-exponential norm as $\norm{x}_{\Psi_1}=\sup_{p\geq 1}\left(\mathbb{E}\abs{x}^p\right)^{1/p}/p$. For a random vector $\bx\in\mathbb{R}^d$, we denote its sub-Gaussian norm as $\norm{\bx}_{\Psi_2}=\sup_{\bv\in\cS^d}\norm{\bv^T\bx}_{\Psi_2}$ and its sub-exponential norm as $\norm{\bx}_{\Psi_1}=\sup_{\bv\in\cS^d}\norm{\bv^T\bx}_{\Psi_1}$. We use $\be_j$ to denote a vector whose elements are all 0 except that the $j$th one is 1.  For a matrix $\bX\in\mathbb{R}^{d_1\times d_2}$, we use $\text{vec}(\bX) \in \RR^{d_1d_2}$ to represent the vector that consists of all the elements from $\bX$ column by column. We use $r(\bX)$, $\norm{\bX}_{\infty}$, $\norm{\bX}_{\text{op}}$, $\nnorm{\bX}$ to denote the rank, elementwise max norm, operator norm and nuclear norm of $\bX$ respectively. We call $\{\bX : \norm{\bX-\bY}_{\infty}\leq r\}$ a $L_{\infty}$-ball centered at $\bY$ with radius r for $r>0$. Define $d_1\wedge d_2 := \min(d_1,d_2)$ and $d_1 \vee d_2 := \max(d_1, d_2)$. For matrices $\bA$ and $\bB$, let $\inn{\bA,\bB}=\text{tr}(\bA ^T\bB)$. For any subspace $\cM \subset \RR^{d \times d}$, define its orthogonal space $\cM^{\perp} := \{\bA: \forall \bM \in \cM, \inn{\bA, \bM} = 0\}$.

\subsection{General Theory}

In this section, we provide a general theorem on the statistical rate of $\widehat\bTheta$ in \eqref{eq:1.7}. As we shall see, the statistical consistency of $\widehat \bTheta$ essentially requires two conditions: i) sufficient penalization $\lambda$; ii) localized restricted strong convexity of $\cL(\bTheta)$ around $\bTheta^*$. In high-dimensional statistics, it is well known that the restricted strong convexity (RSC) of the loss function underpins the statistical rate of the M-estimator \citep{NRW11, RWY10}. In generalized trace regression, however, the fact that the Hessian matrix $\widehat\bH(\bTheta)$ depends on $\bTheta$ creates technical difficulty for verifying RSC for the loss function. To address this issue, we apply the localized analysis due to \cite{FLS15}, where they only require local RSC (LRSC) of $\cL(\bTheta)$ around $\bTheta^*$ to derive statistical rates of $\widehat \bTheta$. Below we formulate the concept of LRSC. For simplicity, from now on we assume that $\bTheta^*$ is a $d$-by-$d$ square matrix. We can easily extend our analysis to the case of rectangular $\bTheta^* \in \RR^{d_1\times d_2}$; the only change in the result is a replacement of $d$ with $\max(d_1, d_2)$ in the statistical rate.

\begin{dfn}
Given a constraint set $\mathcal{C} \subset \mathbb{R}^{d \times d}$, a local neighborhood $\cN$ of $\bTheta^*$, a positive constants $\kappa_{\ell}$ and a tolerance term $\tau_{\ell}$, we say that the loss function $\cL(\cdot)$ satisfies LRSC$(\cC, \cN, \kappa_{\ell}, \tau_{\ell})$ if for all $\bDelta\in\mathcal{C}$ and $\bTheta \in \cN$,
\begin{equation}\label{eq:RSC}
\cL(\bTheta+\bDelta)-\cL(\bTheta)-\inn{\nabla \cL(\bTheta),\bDelta}\geq\kappa_{\ell} \norm{\bDelta}_F^2 - \tau_{\ell}.
\end{equation}
\end{dfn}
Note that $\tau_{\ell}$ is a tolerance term that will be specified in the main theorem. Now we introduce the constraint set $\cC$ in our context. Let $\bTheta^*=\bU\bD\bV^T$ be the SVD of $\bTheta^*$, where the diagonals of $\bD$ are in the decreasing order. Denote the first $r$ columns of $\bU$ and $\bV$ by $\bU^r$ and $\bV^r$ respectively, and define
	\beq
		\label{eq:2.2}
		\begin{aligned}
			& \cM:=\{\bTheta\in \RR^{d \times d}\ |\ \text{row}(\bTheta)\subseteq \text{col}(\bV^r), \text{col}(\bTheta)\subseteq \text{col}(\bU^r)\}, \\
			& \overline{\cM}^{\perp}:=\{\bTheta\in \RR^{d \times d}\ |\ \text{row}(\bTheta)\perp \text{col}(\bV^r), \text{col}(\bTheta)\perp \text{col}(\bU^r)\},
		\end{aligned}
	\eeq
	where $\text{col}(\cdot)$ and $\text{row}(\cdot)$ denote the column space and row space respectively. For any $\bDelta\in \RR^{d \times d}$ and Hilbert space $\cW\subseteq\RR^{d \times d}$, let $\bDelta_{\cW}$ be the projection of $\bDelta$ onto $\cW$. We first clarify here what $\bDelta_\cM$, $\bDelta_{\overline\cM}$ and $\bDelta_{\overline\cM^{\perp}}$ are. Write $\bDelta$ as
	\[
		\bDelta=[\bU^r, \bU^{r^\perp}]\left[
			\begin{array}{cc}
				\bGamma_{11} & \bGamma_{12} \\
				\bGamma_{21} & \bGamma_{22}
			\end{array} \right] [\bV^r, \bV^{r^\perp}]^T,
	\]
	 then the following equalities hold:
	 \beq
	 	\begin{aligned}
			\bDelta_{\cM} =\bU^r\bGamma_{11}{(\bV^r)}^T,\quad\bDelta_{\overline\cM^\perp} = \bU^{r^\perp} \bGamma_{22}(\bV^{r^\perp})^{T}, \quad \bDelta_{\overline\cM} = [\bU^r, \bU^{r^\perp}]\left[
			\begin{array}{cc}
				\bGamma_{11} & \bGamma_{12} \\
				\bGamma_{21} & \bzero
			\end{array} \right] [\bV^r, \bV^{r^\perp}]^T.
		\end{aligned}
	 \eeq
According to Negahban et al. (2012), when $\lambda\geq 2\opnorm{n^{-1}\sum\limits_{i=1}^n \left[b'(\inn{\bX_i, \bTheta^*}) - Y_i\right]\cdot\bX_i}$, regardless of what $r$ is, $\hat{\bDelta}$ falls in the following cone:
 \[
 \mathcal{C}(\mathcal{M},\overline{\mathcal{M}}^{\perp},\bTheta^*):=\Bigl\{ \bDelta \in \RR^{d \times d}: \nnorm{\bDelta_{\overline{\mathcal{M}}^{\perp}}}\leq 3\norm{\bDelta_{\overline{\mathcal{M}}}}_N+4\sum\limits_{j\geq r+1}\sigma_j(\bTheta^*)\Bigr\}.
 \]
	
Now we present the main theorem that serves as a roadmap to establish the statistical rate of convergence for $\widehat \bTheta$.
\begin{thm}\label{thm:1}
Suppose $\mathcal{B}_q(\bTheta^*)\leq\rho$ and
\beq
	\label{eq:lambda}
	\lambda\geq 2\opnorm{\frac{1}{n} \sum\limits_{i=1}^n \left[b'(\inn{\bX_i, \bTheta^*}) - Y_i\right]\cdot\bX_i}.
\eeq
Define $\cN := \{\bTheta\in \RR^{d \times d}: \fnorm{\bTheta - \bTheta^*}^2 \le C_1\rho \lambda^{2-q}, \bTheta - \bTheta^* \in \cC(\cM, \overline\cM^{\perp}, \bTheta^*)\}$ for some constant $C_1$ and let $\tau_{\ell} = C_0\rho \lambda^{2-q}$ for some constant $C_0$.
Suppose $\cL(\bTheta)$ satisfies LRSC$(\cC(\cM, \overline\cM^{\perp}, \bTheta^*), \cN, \allowbreak \kappa_{\ell}, \tau_{\ell})$, where $\cM$ and $\overline\cM$ are constructed as per \eqref{eq:2.2} and $\kappa_{\ell}$ is a positive constant.
Then it holds that
\begin{equation}
	\label{eq:3.2}
\fnorm{\hat{\bDelta}}^2\leq C_1\rho\left(\frac{\lambda}{\kappa_{\ell}}\right)^{2-q}\quad\text{and}\quad\nnorm{\hat{\bDelta}} \leq C_2\rho\left(\frac{\lambda}{\kappa_{\ell}}\right)^{1-q},
\end{equation}
where $C_1, C_2$ are constants.
\end{thm}

Theorem \ref{thm:1} points out two conditions that lead to the statistical rate of $\widehat\bTheta$. First, we need $\lambda$ to be sufficiently large, which has an adverse impact on the rates. Therefore, the optimal choice of $\lambda$ is the lower bound given in \eqref{eq:lambda}. The second requirement is LRSC of $\cL(\bTheta)$ around $\bTheta^*$. In the sequel, for each problem setup we will first derive the rate of the lower bound of $\lambda$ as shown in \eqref{eq:lambda} and then verify LRSC of $\cL(\bTheta)$ so that we can establish the statistical rate.

For notational convenience, later on when we refer to certain quantities as constants, we mean they are independent of $n, d, \rho$. In the next subsections, we will apply the general theorem to analyze various specific problem setups and derive the explicit rates of convergence.

\subsection{Generalized Matrix Regression}\label{sec0}

Generalized matrix regression can be regarded as a generalized linear model (GLM) with matrix covariates. Here we assume that $\vec(\bX_i)$, the vectorized version of $\bX_i$, is a sub-Gaussian random vector with bounded $\psi_2$-norm. Consider $\widehat\bTheta$ as defined in \eqref{eq:1.7}. To derive statistical rate of $\widehat \bTheta$, we first establish the rate of the lower bound of $\lambda$ as characterized in \eqref{eq:lambda}.

\begin{lem}
\label{lem:1}
Consider the following conditions:
\begin{itemize} \itemsep -0.05in
  \item [(C1)] $\{\text{vec}(\bX_i)\}_{i=1}^n$ are i.i.d. sub-Gaussian vectors with $\norm{\text{vec}(\bX_i)}_{\psi_2}\leq\kappa_0<\infty$;
  \item [(C2)] $\abs{b''(x)}\leq M< \infty$ for any $x\in\mathbb{R}$;
 \end{itemize}
Then for any $\nu>0$, there exists a constant $\gamma>0$ such that as long as $d/n<\gamma$, it holds that
\begin{equation}
\mathbb{P}\left(\opnorm{\frac{1}{n}\sum_{i=1}^n(b'(\inn{\bTheta^*,\bX_i})-Y_i) \cdot \bX_i}>\nu\sqrt{\frac{d}{n}}\right)\leq C\exp(-cd),
\end{equation}
where $C$ and $c$ are constants.
\end{lem}

Next we verify the LRSC of $\cL(\bTheta)$.

\begin{lem}\label{lem:2}
Besides (C1) and (C2) in Lemma \ref{lem:1}, assume that
\begin{itemize} \itemsep -0.05in
  \item [(C3)] $\lambda_{min}\left(\bH(\bTheta^*)\right)\geq\kappa_{\ell}>0$;
    \item [(C4)] $\norm{\bTheta^*}_F\geq\alpha\sqrt{d}$ for some constant $\alpha$;
\item [(C5)] $\abs{b'''(x)}\leq {\abs{x}}^{-1}$ for $\abs{x}>1$.
  \end{itemize}
Suppose $\lambda \ge \nu \sqrt{d / n}$, where $\nu$ is the same as in Lemma \ref{lem:1}. Let $\cN = \{\bTheta\in \RR^{d \times d}: \fnorm{\bTheta - \bTheta^*}^2 \le C_1\rho\lambda^{2-q}, \bTheta - \bTheta^* \in \cC(\cM, \overline\cM^{\perp}, \bTheta^*)\}$. As long as $\rho\lambda^{1-q}$ is sufficiently small, $\cL(\bTheta)$ satisfies LRSC$(\cC(\cM, \overline\cM^{\perp}, \bTheta^*), \allowbreak \cN, \kappa, \tau_{\ell})$ with probability at least $1-C_2\exp{(-c_1d)}$, where $\tau_{\ell} = C_0\rho\lambda^{2-q}$, $0<\kappa<\kappa_{\ell}$ and $c_1, C_0,C_1$ and $C_2$ are constants.
\end{lem}

\begin{rmk}
Condition (C4) is mild and is satisfied if there are at least $d$ elements of $\bTheta^*$ that are $\Omega(1)$. Condition (C5) requires that the third order derivative of $b(\cdot)$ decays sufficiently fast.
In fact, except for Poisson regression, most members in the family of generalized linear models satisfy this condition, e.g., linear model, logistic regression, log-linear model, etc.
\end{rmk}

Based on the above two lemmas, we apply Theorem \ref{thm:1} and establish the explicit statistical rate of $\hat{\bTheta}$ as follows.

\begin{thm}\label{thm:2}
Under the conditions in Lemmas \ref{lem:1} and \ref{lem:2}, choosing $\lambda= 2\nu\sqrt{d/n}$, where $\nu$ is the same as in Lemma \ref{lem:1}, there exist constants $\{c_i\}_{i=1}^2$ and $\{C_i\}_{i=1}^5$ such that once $\rho(d / n)^{(1-q) / 2} \le C_1$, we have
\begin{equation}\label{eq:3.11}
\fnorm{\hat{\bTheta}-\bTheta^*}^2\leq C_2\rho\left(\frac{d}{n}\right)^{1-q/2},\quad\nnorm{\hat{\bTheta}-\bTheta^*} \leq C_3\rho\left(\frac{d}{n}\right)^{(1-q)/2}
\end{equation}
with probability at least $1-C_4\exp{(-c_1d)}-C_5\exp{(-c_2d)}$.
\end{thm}

When $q = 0$, $\rho$ becomes the rank of $\bTheta^*$ and there are $O(\rho d)$ free parameters.  Each of these parameters can be estimated at rate $O_P(1/\sqrt{n})$.  Therefore, the sum of squared errors should at least be $O(\rho d / n)$.  This is indeed the bound of $\fnorm{\hat{\bTheta}-\bTheta^*}^2$ given by \eqref{eq:3.11}, which depends on the effective dimension $\rho d$ rather than the ambient dimension $d^2$.  The second result of \eqref{eq:3.11} confirms this in the spectral ``$L_1$-norm'', the nuclear norm.
\subsection{Generalized Reduced-Rank Regression}\label{sec:1}

Consider the conventional reduced-rank regression model (RRR)
\[
	\by_i = {\bTheta^*} \bx_i + \bepsilon_i,
\]
where $\bx_i\in \RR^{d}$ is the covariate, $\by_i\in\RR^{d}$ is the response, $\bTheta^*\in \RR^{d \times d}$ is a near low-rank coefficient matrix and $\bepsilon_i \in \RR^{d}$ is the noise. Again, we set the number of covariates to be the same as the number of responses purely for simplicity of the presentation.   Note that in each sample there are $d$ responses correspondent to the same covariate vector. RRR aims to reduce the number of regression parameters in multivariate analysis. It was first studied in detail by \cite{And51}, where the author considered multi-response regression with linear constraints on the coefficient matrix and applied this model to obtain points estimation and confidence regions in ``shock models'' in econometrics (\cite{Mar50}). Since then, there has been great amount of literature on RRR in econometrics (\cite{ARe94}, \cite{Gew96}, \cite{KPa06}) and statistics (\cite{Ala75}, \cite{VRe13}, \cite{CDC13}).

Now we generalize the above reduced-rank regression to accommodate various types of dependent variables. For any $1 \le i \le n$ and $1 \le j \le d$, $y_{ij}$ is generated from the following density function.
\beq
	\label{eq:2.8}
	f(y_{ij}; \bx_i, \bTheta^*) =  c(y_{ij}) \exp\Bigl( \frac{y_{ij} \eta_{ij}^* - b(\eta_{ij}^*)}{\phi}\Bigr) = c(y_{ij}) \exp\Bigl( \frac{y_{ij}{\btheta_j^*}^{T} \bx_i  - b({\btheta_j^*}^{T} \bx_i)}{\phi}\Bigr),
\eeq
where $\btheta_j^*$ is the $j$th row of $\bTheta^*$, $\eta^*_{ij} = {\btheta_j^*}^{T} \bx_i$, $c(\cdot)$ and $b(\cdot)$ are known functions.
We further assume that for any $(i_1, j_1) \neq (i_2, j_2)$, $y_{i_1j_1} \independent y_{i_2j_2} $. Note that we can recast this model as a generalized trace regression with $N=nd$ samples: $\{\bX_{(i-1)d+j}=\be_j\bx_{i}^T\in\mathbb{R}^{d\times d}, Y_{(i-1)d+j}=y_{ij}\in\mathbb{R}: 1\leq i\leq n, 1\leq j\leq d\}$. We emphasize here that throughout this paper we will use $(\bx_i, \by_i)$ and $\{(\bX_t, Y_t)\}_{t=(i-1)d + 1}^{id}$ to denote the vector and matrix forms of the $i$th sample in RRR.

According to model \eqref{eq:2.8}, we solve for the nuclear norm regularized M-estimator $\widehat\bTheta$ as follows.
\begin{eqnarray}
\label{eq:minmulti}
\hat{\bTheta} & = & \argmin_{\bTheta \in \RR^{d \times d}} \frac{1}{N}\sum_{i=1}^{n}\sum_{j=1}^{d} \left[b'(\inn{\bTheta,\bX_{(i-1)d+j}})-Y_{(i-1)d+j}\cdot
\inn{\bTheta,\bX_{(i-1)d+j}}\right]+\lambda\norm{\bTheta}_N 
    \nonumber \\
& = & \argmin_{\bTheta \in \RR^{d \times d}} \frac{1}{N}\sum_{i=1}^{n}\sum_{j=1}^{d} \left[b'(\btheta_j^T \bx_i)-y_{ij}\cdot \btheta_j^T \bx_i \right]+\lambda\norm{\bTheta}_N. 
\end{eqnarray}
Under the sub-Gaussian design, we are able to derive the covergence rate of $\widehat\bTheta$ in  RRR with the same tool as what we used in matrix regression. Again, we explicitly derive the rate of the lower bound of $\lambda$ in the following lemma.

\begin{lem}\label{lem:3}
Suppose the following conditions hold:
\begin{itemize} \itemsep -0.05in
  \item [(C1)] $\{\bx_i\}_{i=1}^n$ are i.i.d sub-Gaussian vectors with $\norm{\bx_i}_{\psi_2}\leq\kappa_0<\infty$;
  \item [(C2)] $\abs{b''(\cdot)} \le M < \infty$, $\abs{b'''(\cdot)} \le L < \infty$.
 \end{itemize}
Then for any $\nu>0$, there exists a constant $\gamma>0$ such that as long as $d / n<\gamma$, it holds that
\beq
	\label{eq:2.10}
 	P\bigl(\opnorm{\frac{1}{N}\sum \limits_{i=1}^N (b'(\inn{\bX_i, \bTheta^*}) - Y_i)\bX_i } \ge d^{-1}\nu\sqrt{\frac{\phi M \kappa_0 d}{n}}\bigr) \le 2\exp(-cd),
\eeq
where $\phi$ is the same as in \eqref{eq:2.8} and $c$ is a constant.
\end{lem}

The following lemma establishes the LRSC of the loss function.

\begin{lem}\label{lem:4}
Besides conditions in Lemma \ref{lem:3}, assume that
\begin{itemize} \itemsep -0.05in
  \item [(C3)] $\lambda_{min}\left(\bH(\bTheta^*)\right)\geq\kappa_{\ell}>0$.
\end{itemize}
Choose $\lambda = d^{-1} \nu\sqrt{\phi M \kappa_0d / n}$ as in \eqref{eq:2.10}. Let $\cN :=\{\bTheta : \fnorm{\bTheta - \bTheta^*}^ 2 \le \rho\lambda^{2 - q}\}$. For any $\delta>0$, there exists $\gamma>0$ such that when $\rho(d / n)^{1 - q/2} \log(nd)<\gamma$, $\cL(\bTheta)$ satisfies LRSC$(\RR^{d \times d}, \cN, \kappa_{\ell} / (2d), 0)$ with probability at least $1 - 2(nd)^{2 - \frac{\delta}{2}}$.
\end{lem}

Combining to the above lemmas with Theorem 1, we can derive the statistical rate of $\widehat\bTheta$ as defined in \eqref{eq:minmulti}.

\begin{thm}\label{thm:6}
Suppose conditions in Lemmas \ref{lem:3} and \ref{lem:4} hold. Take $\lambda_N =d^{-1} \nu\sqrt{\phi M \kappa_0 d/n}$. For any $\delta > 4$, there exist constants $\{c_i\}_{i=1}^2$ and $\{C_i\}_{i=1}^2$
 such that once $\rho(d / n)^{1 - q/2} \log(nd) <c_1$, any solution to (\ref{eq:minmulti}) satisfies
\begin{equation}
\norm{\hat{\bTheta}-\bTheta^*}_F^2\leq C_1\rho\left(\frac{d}{n}\right)^{1-q/2},\quad\norm{\hat{\bTheta}-\bTheta^*}_N\leq C_2\rho\left(\frac{d}{n}\right)^{(1-q)/2}
\end{equation}
with probability at least $1- 2\exp(-c_2d) - 2(nd)^{2 - \frac{\delta}{2}}$.
\end{thm}

Again, as remarked at the end of Section \ref{sec0}, the error depends on the effective dimension $\rho d$ rather than the ambient dimension $d^2$ for the case $q = 0$.

\subsection{One-Bit Matrix Completion}
Another important example of the generalized trace regression  is the one-bit matrix completion problem, which appears frequently in the online item response questionnaire and recommendation system. The showcase example is the aforementioned Kiva platform in P2P microfinance, in which we only observe sparse binary entries of lenders and borrowers.
Suppose that we have $d_1$ users that answer a small fraction of $d_2$ binary questions. For simplicity of presentation, we again assume that $d_1=d_2 = d$. Specifically, consider the following logistic regression model with $\bX_i=\be_{a(i)}\be_{b(i)}^T\in\mathbb{R}^{d\times d}$.  Namely, the $i$th data records the $a(i)$th user answers the binary question $b(i)$.  The problem is also very similar to the aforementioned Netflix problem, except that only dichotomous responses are recorded here.

The logistic regression model assumes that
\begin{equation}
\label{eq:2.1}
\log{\frac{\mathbb{P}\left(Y_i=1\given  \bX_i \right)}{\mathbb{P}\left(Y_i=0 \given  \bX_i \right)}}=\text{tr}(\bTheta^{*T}\bX_i)=\Theta^*_{a(i), b(i)}.
\end{equation}
Note that this model can be derived from generalized trace regression \eqref{eq:model} with $b'(\eta^*_i) = (1 + \exp(-\eta^*_i))^{-1}$. \eqref{eq:2.1} says that given $\bX_i = \be_{a(i)}\be_{b(i)}^T\in\mathbb{R}^{d\times d}$, $Y_i$ is a Bernoulli random variable with $\bbP(Y_i = 1 \given  \bX_i) = (1 + \exp(- \Theta^*_{a(i), b(i)}))^{-1}$. We assume that $\{(a(i), b(i))\}_{i=1}^N$ are randomly and uniformly distributed over $\{(j, k)\}_{1 \le j \le d, 1 \le k \le d}$. We further require $\bTheta^*$ to be non-spiky in the sense that $\supnorm{\bTheta^*}=O(1)$ and thus $\fnorm{\bTheta^*} = O(d)$. This condition ensures consistent estimation as elucidated in \cite{NWa12}. For ease of theoretical reasoning, from now on we will rescale the design matrix $\bX_i$ and the signal $\bTheta^*$ such that $\bX_i = d\be_{a(i)}\be_{b(i)}^T$ and $\norm{\bTheta^*}_F\leq 1$. Based on such setting, we estimate $\bTheta^*$ through minimizing negative log-likelihood plus nuclear norm penalty under a elementwise max-norm constraint:
\begin{equation}\label{eq:mincompletion}
\hat{\bTheta}=\argmin_{\norm{\bTheta}_{\infty}\leq R/ d} \left\{\frac{1}{n}\sum_{i=1}^n \left[\log (1+\exp({\inn{\bTheta,\bX_i}}))-Y_i\inn{\bTheta,\bX_i}\right]+\lambda\nnorm{\bTheta}\right\},
\end{equation}
where $\lambda$ and $R$ are tuning parameters.

Again, we first derive the rate of the lower bound for $\lambda$ as shown in Theorem 1. For this specific model, simple calculation shows that the lower bound \eqref{eq:lambda} reduces to
$$
\|n^{-1}\sum_{i=1}^n [\exp(\bTheta^*, \bX_i) / \allowbreak (1 + \exp(\bTheta^*, \bX_i))-Y_i ]\cdot\bX_i\|_{\text{op}}.
$$

\begin{lem}\label{lem:5}
Under the following conditions:
\begin{itemize}

\item [(C1)] $\norm{\bTheta^*}_F\leq 1$, $\norm{\bTheta^*}_{\infty}\leq R/d$ where $0<R<\infty$;
\item [(C2)] $\{\bX_i\}_{i=1}^n$ are uniformly sampled from $\left\{d\be_j\be_k^T\right\}_{1\leq j,k \leq d}$;
\end{itemize}
For any $\delta>1$, there exists $\gamma>0$ such that as long as $d\log{d}/n<\gamma$, the following inequality holds for some constant $\nu>0$:
\begin{equation}
\mathbb{P}\left(\opnorm{\frac{1}{n}\sum\limits_{i=1}^n\bigl(\frac{\exp{(\inn{\bTheta^*,\bX_i)}}}{\exp{(\inn{\bTheta^*,\bX_i})}+1}-Y_i\bigr )\bX_i}>\nu\sqrt{\frac{\delta d\log{d}}{n}}\right)\leq 2d^{1-\delta}.
\end{equation}
\end{lem}

Next we study the LRSC of the loss function. Following \cite{NWa12}, besides $\cC(\cM, \overline \cM ^{\perp}, \bTheta^*)$, we define another constraint set
\begin{equation}\label{eq:25}
\cC'(c_0): =\left\{\bDelta\in\mathbb{R}^{d\times d},\bDelta\neq \bzero : \frac{\norm{\bDelta}_{\infty}}{\norm{\bDelta}_F}\cdot \frac{\norm{\bDelta}_N}{\fnorm{\bDelta}}\leq\frac{1}{c_0d}\sqrt{\frac{n}{d\log{d}}}\right\}.
\end{equation}
Here $\supnorm{\bDelta}/ \fnorm{\bDelta}$ and $\nnorm{\bDelta} / \fnorm{\bDelta}$ are measures of spikiness and low-rankness of $\bDelta$. Let $\cN= \{\bTheta: \supnorm{\bTheta - \bTheta^*} \le 2R/d\}$. Note that $\cN$ is not the same as in Theorem \ref{thm:1} any more. As we shall see later, instead of directly applying Theorem \ref{thm:1}, we need to adapt the proof of Theorem \ref{thm:1} to the matrix completion setting to derive statistical rate of $\widehat \bTheta$. The following lemma establishes LRSC$(\cC'(c_0), \cN, \kappa_{\ell}, 0)$ of $\cL(\bTheta)$ for some $\kappa_{\ell} > 0$.

\begin{lem}\label{lem:6}
There exist constants $C_1, C_2, c_1, c_2$
such that as long as $n>C_1d\log{d}$ and $R \le c_1$, it holds with probability greater than $1-C_2\exp{(-c_2d\log{d})}$ that for all $\bDelta\in\mathcal{C}'(c_0)$ and $\bTheta \in \cN$,

\begin{equation}\label{eq:26}
\vec{({\bDelta})}^T\hat{\bH}({\bTheta})\vec{({\bDelta})}\geq\frac{\norm{{
\bDelta}}_F^2}{512(\exp(R)+\exp(-R)+2)}.
\end{equation}
\end{lem}

Now we are ready to establish the statistical rate of $\widehat\bTheta$ in \eqref{eq:mincompletion}.

\begin{thm}\label{thm:4}
Let $\widehat \bTheta$ be defined by \eqref{eq:mincompletion}. Suppose the conditions (C1) and (C2) in Lemma \ref{lem:5} hold for a sufficiently small $R$ and $\mathcal{B}_q(\bTheta^*)\leq\rho$. Consider any solution $\hat{\bTheta}$ to (\ref{eq:mincompletion}) with parameter $\lambda=2\nu\sqrt{\delta d\log{d}/n}$, where $\delta > 1$. There exist constants $\{C_i\}_{i=0}^4$ such that as long as $n>C_0 d\log{d}$,
\begin{equation}
\begin{split}
&\norm{\hat{\bTheta}-\bTheta^*}_F^2\leq C_1\max\Bigl\{\rho\left(\sqrt{\frac{d\log{d}}{n}}\right)^{2-q},\frac{R^2}{n}\Bigr\} \\
&\norm{\hat{\bTheta}-\bTheta^*}_N\leq C_2\max\Bigl\{\rho\left(\sqrt{\frac{d\log{d}}{n}}\right)^{1-q},\left(\rho\left(\frac{R^2}{n}\right)^{1-q}\right)^{\frac{1}{2-q}}\Bigr\} \\
\end{split}
\end{equation}
with probability at least $1-C_3\exp{(-C_4d\log{d})}-2d^{1-\delta}$.

\end{thm}

\begin{rmk}
In \cite{davenport20141}, they derived that $\norm{\hat{\bTheta}-\bTheta^*}_F^2 = O_P(\sqrt{\rho d/n})$ when $\bTheta^*$ is exactly low-rank.
This is slower than our rate $O_P(\rho d/ n)$. Moreover, we provide an extra bound on the nuclear norm of the error.
\end{rmk}


\section{Simulation Study}\label{sec:3}

\subsection{Generalized Matrix Regression}\label{sec:4.2}

In this section, we verify the statistical rates derived in (\ref{eq:3.11}) through simulations. We let $d=20, 40$ and $60$. For each dimension, we take $n$ to be $1800, 3600, 5400, 7200$ and $9000$. We set $\bTheta^* \in \RR^{d \times d}$ with $r(\bTheta^*)=5$ and all the nonzero singular values of $\bTheta^*$ equal to $1$. Each design matrix $\bX_i$ has i.i.d. entries from $\cN(0,1)$ and $Y_i \sim \text{Bin}(0,\exp(\eta_i^*)/(1+\exp(\eta_i^*)))$, where $\eta_i^*=\inn{\bTheta^*,\bX_i}$. We choose $\lambda \asymp \sqrt{d/n}$ and tune the constant before the rate for optimal performance.

\begin{figure}[htbp]
\centering
\includegraphics[scale=0.4]{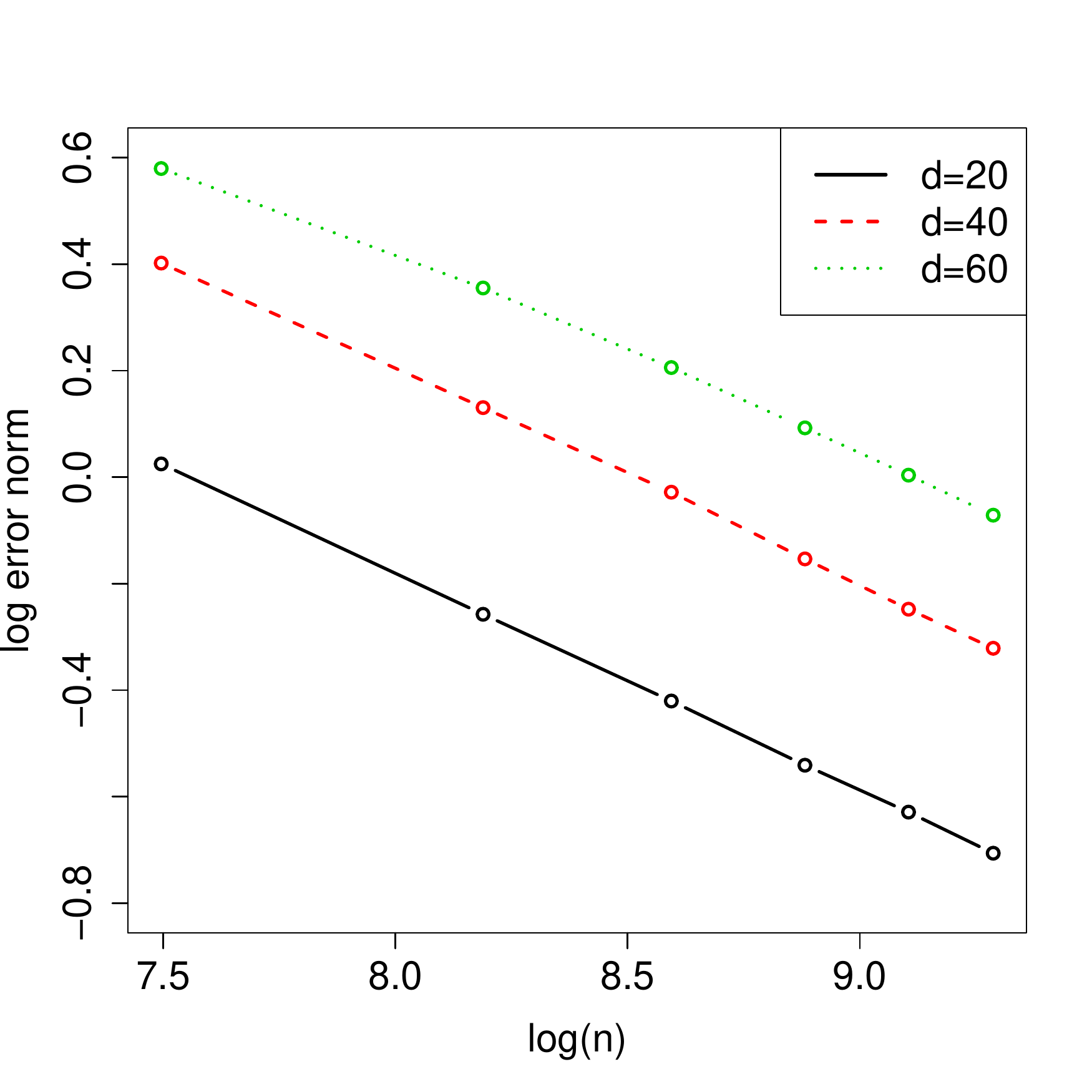}
\caption{$\log\fnorm{\hat{\bTheta}-\bTheta^*}$ versus $\log(n)$ for different dimension $d$.}
\label{fig:mcs}
\end{figure}

Our simulation is based on $100$ independent replications, where we record the estimation error in terms of the logarithmic Frobenius norm $\log \fnorm{\widehat \bTheta - \bTheta^*}$. The averaged statistical error is plotted against the logarithmic sample size in Figure \ref{fig:mcs}.
As we can observe from the plot, the slope of curve is almost $-1/2$, which is consistent with the order of $n$ in the statistical rate we derived for $\widehat\bTheta$.  The intercept also matches the order of $d$ in our theory.  For example, in the plot, the difference between the green and red lines predicted by the theory is $(\log(60)-\log(40))/2=0.20$, which is in line with the empirical plot.  Similarly, the difference between the red and black lines should be around $(\log(40)-\log(20))/2=0.35$, which is also consistent with the plot.

To solve the optimization problem (\ref{eq:1.7}), we exploit an iterative Peaceman-Rachford splitting method. We start from $\widehat\bTheta^{(0)}={\bf 0}$. In the $k$th step, we take the local quadratic approximation of $\cL_n(\bTheta)$ at $\bTheta = \bTheta^{(k - 1)}$:
\begin{equation}
	\label{eq:4.4}
\begin{aligned}
{\cL}^{(k)}_n(\bTheta)=&\frac{1}{2}\vec(\bTheta-\bTheta^{(k-1)})^{T}\nabla^2_{\bTheta}\cL_n(\bTheta^{(k-1)})\vec(\bTheta-\bTheta^{(k-1)})+\inn{\nabla_{\bTheta}\cL_n(\bTheta^{(k-1)}), \bTheta-\bTheta^{(k-1)}} \\
&+\cL_n(\bTheta^{(k-1)}).
\end{aligned}
\end{equation}
and then solve the following optimization problem to obtain $\widehat \bTheta^{(k)}$:
\begin{equation}\label{eq:3.2}
\widehat \bTheta^{(k)}=\argmin_{\bTheta} \cL^{(k)}_n(\bTheta) + \lambda\norm{\bTheta}_N .
\end{equation}
We borrow the algorithm from \cite{fan2016shrinkage} to solve the optimization problem \eqref{eq:3.2}. In Section 5.1 of \cite{fan2016shrinkage}, they applied a contractive Peaceman-Rachford splitting method to solve a nuclear norm penalized least square problem:
\begin{equation}\label{eq:4.3}
\begin{split}
\hat{\bTheta}&=\argmin_{\bTheta}\left\{\frac{1}{n}\sum\limits_{i=1}^n(Y_i-\inn{\bTheta,\bX_i})^2+\lambda\norm{\bTheta}_N\right\} \\
&=\argmin_{\bTheta}\left\{\vec(\bTheta)^{T}\frac{1}{n}\sum\limits_{i=1}^n\vec(\bX_i)\vec(\bX_i)^{T}\vec(\bTheta)+\inn{\frac{2}{n}\sum\limits_{i=1}^nY_i\bX_i,\bTheta}+\lambda\norm{\bTheta}_N\right\}.
\end{split}
\end{equation}
Construct
\[
	\tilde{\bX}_i^{(k)}= \frac{1}{n}\sum\limits_{i=1}^n\sqrt{b''(\inn{\widehat\bTheta^{(k-1)},\bX_i})}\bX_i
\]
and
\[
	\tilde{Y}_i^{(k)}= \frac{1}{n}\sum\limits_{i=1}^n b''(\inn{\widehat\bTheta^{(k-1)},\bX_i})^{-\frac{1}{2}}\allowbreak \left[Y_i-b'(\inn{\widehat\bTheta^{(k-1)},\bX_i})\right].
\]
Some algebra shows  that the following nuclear norm penalized least square problem is equivalent to \eqref{eq:3.2}
\begin{equation}\label{eq:4.4}
\begin{split}
\hat{\bTheta}^{(k)}=\argmin_{\bTheta}{\Big \{}&\frac{1}{2}\vec(\bTheta-\widehat\bTheta^{(k-1)})^{T}\frac{1}{n}\sum\limits_{i=1}^n\vec(\tilde{\bX}_i^{(k)})\vec(\tilde{\bX}_i^{(k)})^T \vec(\bTheta-\widehat\bTheta^{(k-1)}) \\
&+\inn{\frac{1}{n}\sum\limits_{i=1}^n\tilde{Y}_i^{(k)}\tilde{\bX}_i^{(k)},\bTheta-\widehat\bTheta^{(k-1)}}+\lambda\norm{\bTheta}_N{\Big \}}.
\end{split}
\end{equation}
We can further write (\ref{eq:4.4}) as an optimization problem of minimizing the sum of two convex functions:
\begin{equation*}
\begin{aligned}
& \underset{x}{\text{minimize}}
& & \frac{1}{2n}\sum\limits_{i=1}^n\left(\tilde{Y}_i^{(k)}-\inn{\bTheta_x,\tilde{\bX}_i^{(k)}}\right)^2+\lambda\norm{\bTheta_y}_N \\
& \text{subject to}
& & \bTheta_x-\bTheta_y=-\bTheta^{(k-1)}.
\end{aligned}
\end{equation*}
It has been explicitly explained in \cite{fan2016shrinkage} on how to solve the above optimization problem using the Peaceman-Rachford splitting method. We provide the algorithm that is specific to our problem here. Here we first define the singular value soft thresholding operator $\cS_{\tau}(\cdot )$. For any $\bX \in \RR^{d \times d}$, let $\bX = \bU \bD \bV^T$ be its SVD, where $\bU$ and $\bV$ are two orthonormal matrices and $\bD = \diag(\sigma_1, \ldots, \sigma_d)$ with $\sigma_1 \ge \ldots \ge \sigma_d$. Then $\cS_{\tau}(\bX) := \bU \widetilde\bD \bV^T$, where $\widetilde\bD := \diag(\max(\sigma_1 - \tau, 0), \max(\sigma_2 - \tau, 0), \ldots, \max(\sigma_d - \tau, 0))$. Let $\mathbb{X}^{(k)}$ be an $n\times d^2$ matrix whose rows are $\vec(\tilde{\bX}_i^{(k)})$ and $\mathbb{Y}^{(k)}$ be the response vector $\tilde{Y}^{(k)}$. For $\ell=0, 1, \ldots$, 
\beq
                \left\{
                \begin{aligned}
                & \btheta_x^{(\ell+1)}=(2\mathbb{X}^{(k)\top}\mathbb{X}^{(k)}/n+ \beta \cdot \bI )^{-1}(\beta\cdot (\btheta_y^{(\ell)}-\vec(\widehat\bTheta^{(k-1)}))+\brho^{(\ell)} + 2\mathbb{X}^{(k)\top}\mathbb{Y}^{(k)}/n), \\
                & \brho^{(\ell + \frac{1}{2})}= \brho^{(\ell)}-\alpha\beta(\btheta_x^{(\ell + 1)}-\btheta_y^{(\ell)}+\vec(\widehat\bTheta^{(k-1)})), \\
                & \btheta_y^{(\ell + 1)}=\vec(\cS_{2\lambda/\beta}(\text{mat}(\btheta_x+\vec(\widehat\bTheta^{(k-1)})-\brho^{(\ell + \frac{1}{2})} / \beta))), \\
                & \brho^{(\ell + 1)}= \brho^{(\ell + \frac{1}{2})}-\alpha\beta(\btheta_x^{(\ell + 1)}+\vec(\widehat\bTheta^{(k-1)})-\btheta_y^{(\ell + 1)}),
                \end{aligned}
                \right .
                \label{eq:csiter}
\eeq
where we choose $\alpha=0.9$ and $\beta=1$. $\btheta_x^{(\ell)}, \btheta_y^{(\ell)}\in\mathbb{R}^{d^2}$ for $\ell\geq0$ and we can initialize them by $\btheta_x^{(0)}=\btheta_y^{(0)}={\bf 0}$. When $\btheta_x^{(\ell)}$ and $\btheta_y^{(\ell)}$ converge, we reshape $\btheta^{(\ell)}_y$ as a $d\times d$ matrix and return it as $\widehat\bTheta^{(k)}$. We iterate this procedure until $\fnorm{\widehat\bTheta^{(k)}-\widehat\bTheta^{(k-1)}}$ is smaller than $10^{-3}$ and return $\widehat\bTheta^{(k)}$ as the final estimator of $\bTheta^*$.

\subsection{Generalized Reduced-Rank Regression}

\begin{figure}[htbp]
\centering
\includegraphics[scale=0.4]{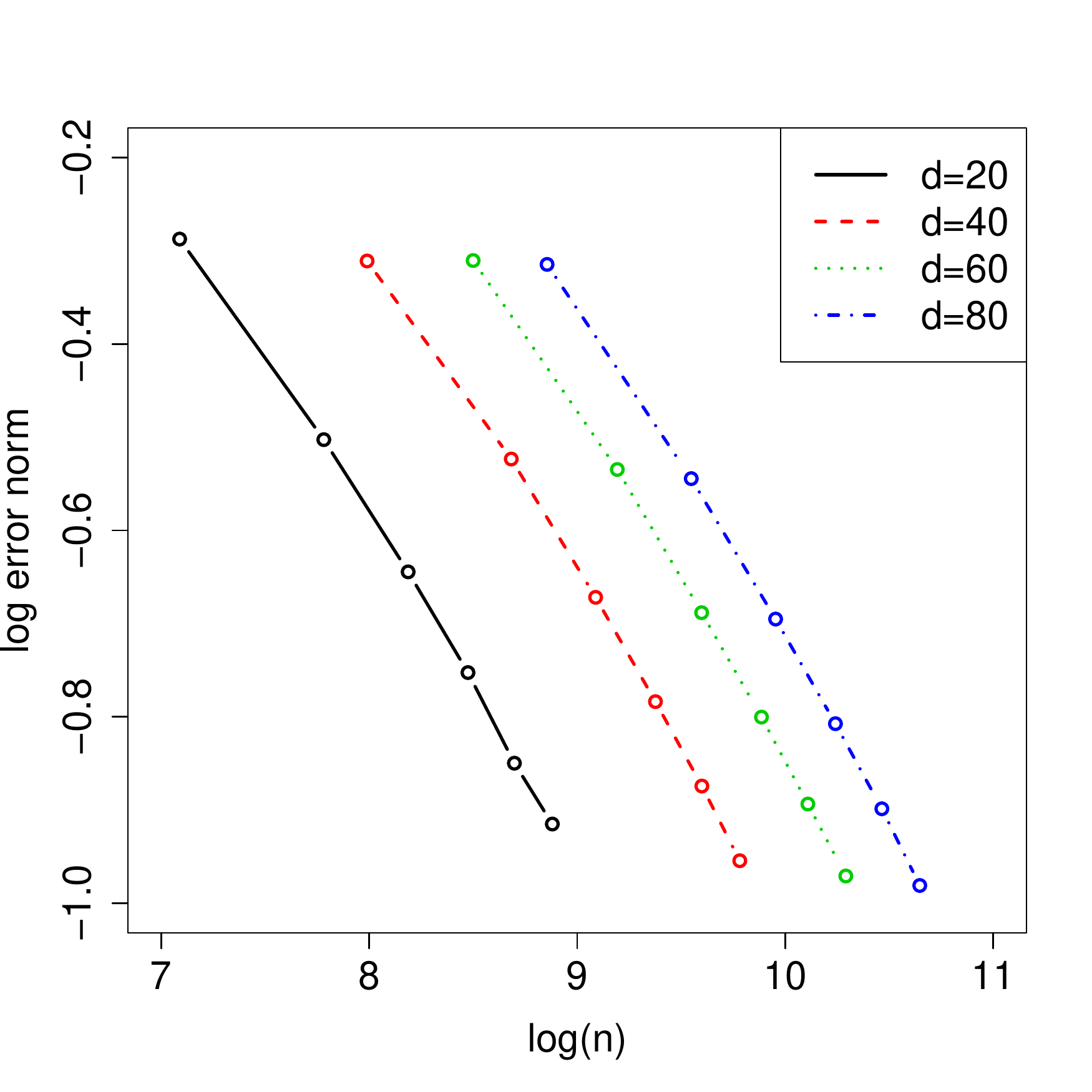}
\includegraphics[scale=0.4]{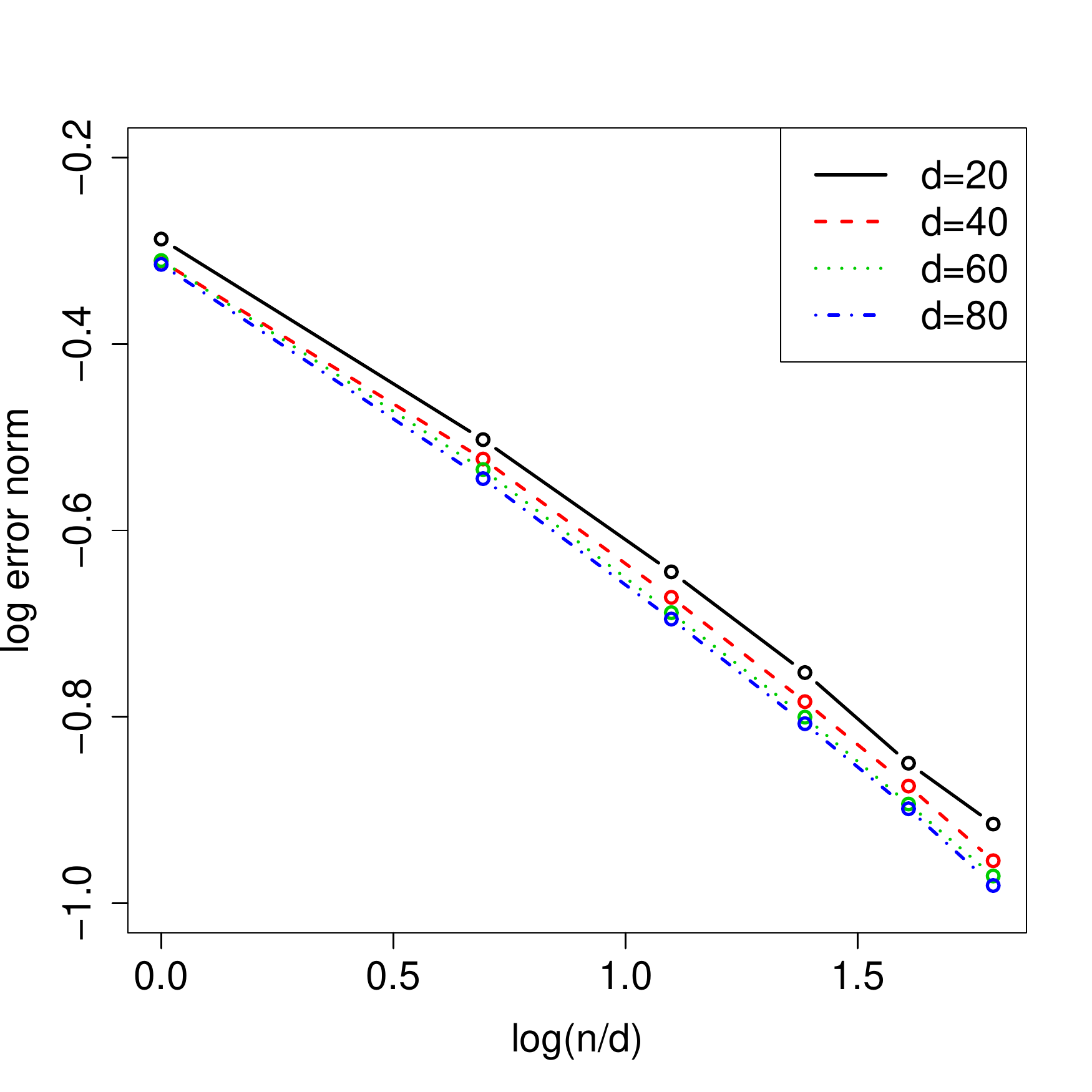}
\caption{$\log\fnorm{\hat{\bTheta}-\bTheta^*}$ versus $\log(n)$ and log standardized sample size $\log\left(n/d\log(d)\right)$.}
\label{fig:multi}
\end{figure}

In this section, we let $d=20, 40, 60$ and $80$. For each dimension, we consider 6 different values for $n$ such that $n / (d\log{(d)}) =20, 40, 60, 80, 100$ and $120$. We set the rank of $\bTheta^*$ to be $5$ and let $\fnorm{\bTheta^*}=1$. For $1\leq i\leq n$ and $1\leq j\leq d$, we let the covariate $\bx_i$ have i.i.d. entries from $\cN(0,1)$ and let $y_{ij}$ follow $\text{Bin}(0,\exp(\eta^*)/(1+\exp(\eta^*)))$ where $\eta^*=\inn{\bTheta^*_j,\bx_i}$. We choose $\lambda \asymp \sqrt{d \log(d)/n}$ and tune the constant before the rate for optimal performance. The experiment is repeated for 100 times and the logarithmic Frobenius norm of the estimation error is recorded in each repetition. We plot the averaged statistical error in Figure \ref{fig:multi}.

We can see from the left panel that the logarithmic error decays as logarithmic sample size grows and the slope is almost $-1/2$. The right panel illustrates that when we standardize the sample size by $d\log(d)$, the statistical error curves are well-aligned, which is consistent with the statistical error rate in our theorem.

As for the implementation, we again use the iterative Peaceman-Rachford splitting method to solve for the estimator. We start from $\widehat\bTheta^{(0)}={\bf 0}$. In the $k$th step $(k \ge 1)$, let \[\bS^{(k)}=\frac{1}{nd}\sum\limits_{i=1}^n\sum\limits_{j=1}^d \frac{\exp(\inn{\widehat\bTheta^{(k-1)}_j,\bx_i})}{(1+\exp(\inn{\widehat\bTheta^{(k-1)}_j,\bx_i}))^2}\bx_i\bx_i^{T},\]
\[\tilde{y}_{ij}^{(k)}=y_{ij}-\frac{\exp(\inn{\widehat\bTheta^{(k-1)}_j,\bx_i})}{1+\exp(\inn{\widehat\bTheta^{(k-1)}_j,\bx_i})}\quad\text{and}\quad \bT^{(k)}=\sum\limits_{i=1}^n\bx_i\tilde{\by}_i^{T}.\]
We iterate the following algorithm to solve for $\widehat\bTheta^{(k)}$. Here $\alpha=0.9$ and $\beta=1$.
\beq
                \left\{
                \begin{aligned}
                & \bTheta_x^{(\ell + 1)}=(2\bS^{(k)}/n+ \beta \cdot \bI )^{-1}(\beta\cdot (\bTheta_y^{(\ell)}-\widehat\bTheta^{(k-1)})+\brho^{(\ell)} + 2\bT^{(k)}/n), \\
                & \brho^{(\ell + \frac{1}{2})}= \brho^{(\ell)}-\alpha\beta(\bTheta_x^{(\ell + 1)}+\widehat\bTheta^{(k-1)}-\bTheta_y^{(\ell)}), \\
                & \bTheta_y^{(\ell + 1)}=\cS_{2\lambda/\beta}(\bTheta_x+\widehat\bTheta^{(k-1)}-\brho^{(\ell + \frac{1}{2})} / \beta), \\
                & \brho^{(\ell + 1)}= \brho^{(\ell + \frac{1}{2})}-\alpha\beta(\bTheta_x^{(\ell + 1)}+\widehat\bTheta^{(k-1)}-\bTheta_y^{(\ell + 1)}).
                \end{aligned}
                \right.
  \eeq
Here, $\cS_{\tau}(\cdot)$ is the singular value soft thresholding function we introduced in Section \ref{sec:4.2}. Note that $\bTheta_x^{(\ell)}, \bTheta_y^{(\ell)}\in\mathbb{R}^{d\times d}$ for all $\ell\geq 0$ and they are irrelevant to $\widehat\bTheta^{(k)}$ though they share similar notations. We start from $\bTheta_x^{(0)}=\bTheta_y^{(0)}={\bf 0}$ and iterate this procedure until they converge. We return the last $\bTheta_y^{(\ell)}$ to be $\widehat\bTheta^{(k)}$.

We repeat the above algorithm until $\fnorm{\widehat\bTheta^{(k)}-\widehat\bTheta^{(k-1)}}$ is smaller than $10^{-3}$ and take $\widehat\bTheta^{(k)}$ as the final estimator of $\bTheta^*$.

\subsection{1-Bit Matrix Completion}

\subsubsection{Statistical consistency}

\begin{figure}
\centering
\includegraphics[scale=0.4]{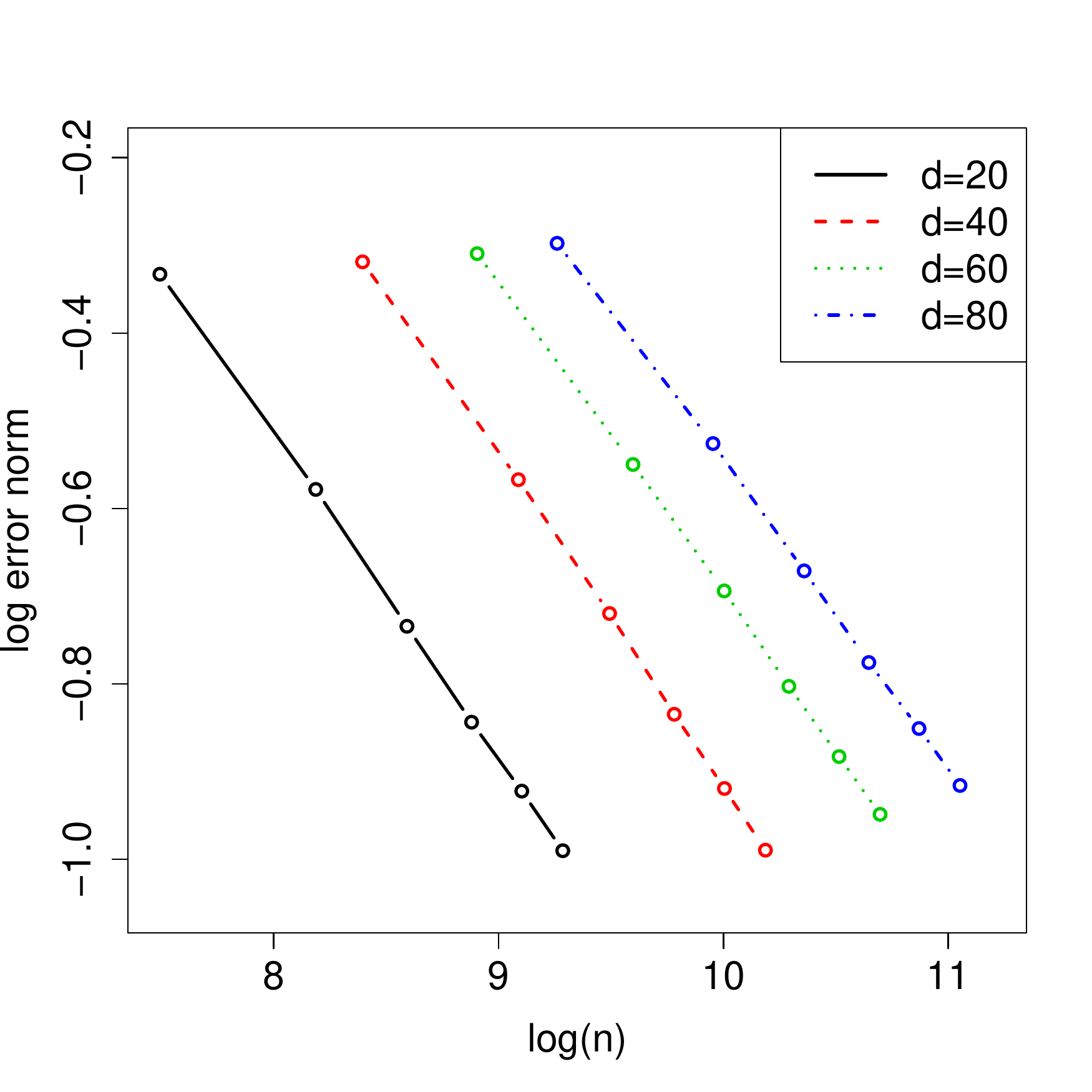}
\includegraphics[scale=0.4]{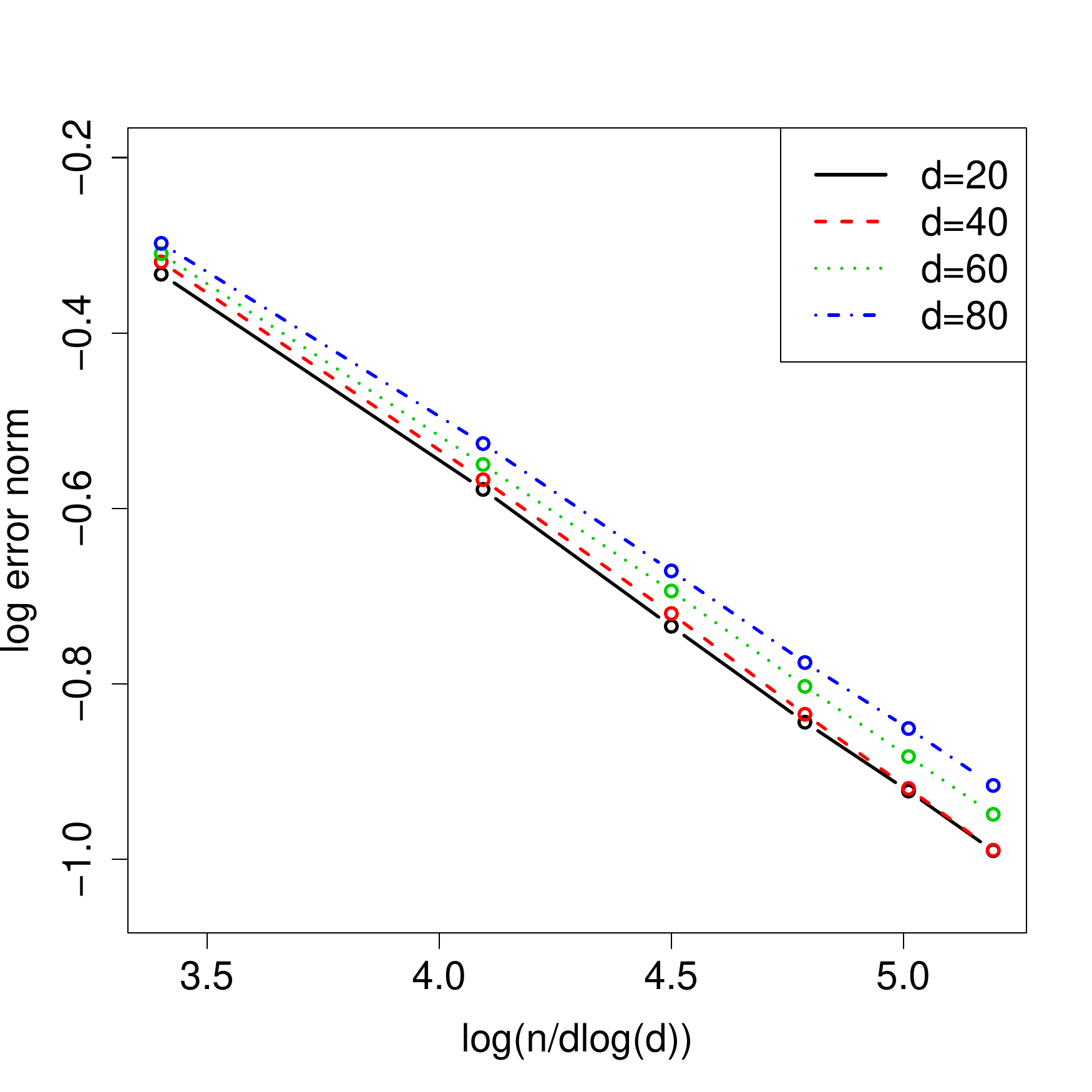}
\caption{$\log\fnorm{\hat{\bTheta}-\bTheta^*}$ versus $\log n$ and $\log (n / d )$.}  
\label{fig:1}
\end{figure}

We consider $\bTheta^* \in \RR^{d \times d}$ with dimension $d=20, 40, 60$ and $80$. For each dimension, we consider 6 different values for $n$ such that $n / (d\log{d}) =30, 60, 90, 120, 150$ and $180$. We let $\text{r}(\bTheta^*)=5$, $\fnorm{\bTheta^*}=1$ and $R=2\norm{\bTheta^*}_{\infty}$. The design matrix $\bX_i$ is a singleton and it is uniformly sampled from $\{\be_j\be_k^{T}\}_{1\leq j,k\leq d}$. We choose $\lambda \asymp \sqrt{d \log(d)/n}$ and tune the constant before the rate for optimal performance. The experiment is repeated for 100 times and the logarithmic Frobenius norm of the estimation error is recorded in each repetition. We plot the averaged statistical error against the logarithmic sample size in Figure \ref{fig:1}.





We can see from the left panel in Figure \ref{fig:1} that $\log \fnorm{\widehat \bTheta - \bTheta^*}$ decays as $\log n$ grows and the slope is almost $-1/2$. Meanwhile, Theorem \ref{thm:4} says that $\log \fnorm{\widehat \bTheta - \bTheta^*}$ should be proportional to $\log(d\log{d}/n)$. The right panel of Figure \ref{fig:1} verifies this rate: it shows that the statistical error curves for different dimensions are well-aligned if we adjust the sample size to be $n/d\log{d}$.

To solve the optimization problem in (\ref{eq:mincompletion}), we exploit the ADMM method used in Section 5.2 in \cite{fan2016shrinkage}. In \cite{fan2016shrinkage}, they minimized a quadratic loss function with a nuclear norm penalty under elementwise max norm constraint. Our goal is to replace the quadratic loss therein with negative log-likelihood and solve the optimization problem. Here we iteratively call the ADMM method in \cite{fan2016shrinkage} to solve a series of optimization problems whose loss function is local quadratic approximation of the negative log-likelihood. We initialize $\bTheta$ with $\hat{\bTheta}^{(0)}={\bf 0}$ and introduce the algorithm below.

In the $k$th step, we take the local quadratic approximation of $\cL_n(\bTheta)$ at $\bTheta = \widehat\bTheta^{(k - 1)}$:
\begin{equation}
	\label{eq:4.0}
\begin{aligned}
{\cL}^{(k)}_n(\bTheta)=&\frac{1}{2}\vec(\bTheta-\widehat\bTheta^{(k-1)})^{T}\nabla^2_{\bTheta}\cL_n(\widehat\bTheta^{(k-1)})\vec(\bTheta-\widehat\bTheta^{(k-1)})+\inn{\nabla_{\bTheta}\cL_n(\widehat\bTheta^{(k-1)}), \bTheta-\widehat\bTheta^{(k-1)}} \\
&+\cL_n(\widehat\bTheta^{(k-1)}).
\end{aligned}
\end{equation}
and solve the following optimization problem to obtain $\widehat\bTheta^{(k)}$:
\begin{equation}\label{eq:4.2}
\widehat \bTheta^{(k)}=\argmin_{\bTheta} \cL^{(k)}_n(\bTheta) + \lambda\norm{\bTheta}_N .
\end{equation}
To solve the above optimization problem, we borrow the algorithm proposed in \cite{fang2015max}. Let $\bL, \bR, \bW\in\mathbb{R}^{2d\times 2d}$ be the variables in our algorithm and let $\bL^{(0)}=\bR^{(0)}={\bf 0}$. Define
\[\bTheta_{jk}^a=\sum\limits_{i=1}^n\frac{\exp(\inn{\bTheta,\bX_i})}{(1+\exp(\inn{\bTheta,\bX_i}))^2}\mathds{1}_{\{\bX_i = \be_j\be_k^{T}\}},\]
\[\bTheta_{jk}^b=\sum\limits_{i=1}^n\left[Y_i-\frac{\exp(\inn{\bTheta,\bX_i})}{1+\exp(\inn{\bTheta,\bX_i})}\right]\mathds{1}_{\{\bX_i = \be_j\be_k^{T}\}}.\]
We introduce the algorithms of the variables in our problem and interested readers can refer to \cite{fang2015max} for the technical details in the derivation and stopping criteria of the algorithm. For $\ell \ge 1$,
\beq
                \left\{
                \begin{aligned}
&\bL^{(\ell+1)}=\Pi_{\cS_+^{2d}}\left\{\bR^{(\ell)}+\left(\begin{aligned}{\bf 0} \quad\quad& \widehat\bTheta^{(k-1)} \\ \widehat\bTheta^{(k-1)} \quad& \quad{\bf 0} \\\end{aligned}\right)-\rho^{-1}(\bW^{(\ell)}+2\lambda\bI)\right\}=\left(\begin{aligned}{[\bL^{(\ell+1)}]^{11}} \quad& [\bL^{(\ell+1)}]^{12} \\ [\bL^{(\ell+1)}]^{21} \quad& [\bL^{(\ell+1)}]^{22} \\\end{aligned}\right),\\
&\bC= \left(\begin{aligned}\bC^{11} \quad& \bC^{12} \\ \bC^{21} \quad& \bC^{22} \\\end{aligned}\right)=\bL^{(\ell+1)}-\left(\begin{aligned}{\bf 0} \quad\quad& \widehat\bTheta^{(k-1)} \\ \widehat\bTheta^{(k-1)} \quad& \quad{\bf 0} \\\end{aligned}\right)+\bW^{(\ell)}/\rho,\\
&\bR_{jk}^{12}=\Pi_{[-R,R]}\left\{(\rho\bC_{jk}^{12}+2\bTheta_{jk}^b/n)/(\rho+2\bTheta_{jk}^a/n)\right\}, 1\leq j\leq d, 1\leq k\leq d, \\
&\bR^{(\ell+1)}= \left(\begin{aligned}\bC^{11}\quad& \bR^{(12)} \\ (\bR^{12})^{T} \quad& \bC^{22} \\\end{aligned}\right),\\
&\bW^{(\ell+1)}=\bW^{(\ell)}+\gamma\rho(\bL^{(\ell+1)}-\bR^{(\ell+1)}-\left(\begin{aligned}{\bf 0} \quad\quad& \widehat\bTheta^{(k-1)} \\ \widehat\bTheta^{(k-1)} \quad& \quad{\bf 0} \\\end{aligned}\right)).
                \end{aligned}
                \right .
                \label{eq:csiter}
\eeq
In the algorithm, $\Pi_{\cS_+^{2d}}(\cdot)$ represents the projection operator onto the space of positive semidefinite matrices $\cS_+^{2d}$, $\rho$ is taken to be 0.1 and $\gamma$ is the step length which is set to be 1.618. When the algorithm converges and stops, we elementwise truncate $\bL^{12}$ at the level of $R$ and return the truncated $\tilde{\bL}^{12}$ as $\widehat\bTheta^{(k)}$. Specifically, $\tilde{\bL}_{jk}^{12}=\sgn(\bL_{jk}^{12}) ( | \bL_{jk}^{12} |\wedge R )$ for $1\leq j\leq d, 1\leq k\leq d$.

When $\fnorm{\widehat\bTheta^{(k)}-\widehat\bTheta^{(k-1)}}$ is smaller than $10^{-3}$, we return $\widehat\bTheta^{(k)}$ as our final estimator of $\bTheta^*$.

\subsubsection{Comparison between GLM and linear model}

As we mentioned in the introduction, the motivation of generalizing trace regression is to accommodate the dichotomous response in recommending systems such as Netflix Challenge, Kiva, etc. In this section, we compare the performance of generalized trace regression and standard trace regression in  predicting discrete ratings.

The setting is very similar to the last section. We set $\bTheta^*$ to be a square matrix with dimension $d=20, 40, 60$ and $80$. We let $r(\bTheta^*)=5$ and its top five eigenspace be the top five eigenspace of the sample covariance matrix of $100$ random vectors following $\cN(\bzero, \bI_{d})$. For each dimension, we consider 10 different values for $n$ such that $n/d\log{d}=1,2,...,10$. and generate the true rating matrix $\bT$ in the following way:
\[T_{i,j}=\begin{cases}1\quad\quad\text{w.p.}\quad\frac{\exp(\bTheta^*_{ij})}{1+\exp{(\bTheta^*_{ij})}} \\ 0\quad\quad\text{w.p.}\quad\frac{1}{1+\exp{(\bTheta^*_{ij})}} \\\end{cases}\quad1\leq i \leq d, 1\leq j \leq d.\]
We will show that generalized trace regression outperforms the linear trace regression in prediction.

\begin{figure}[htbp]
\centering
\includegraphics[scale=0.4]{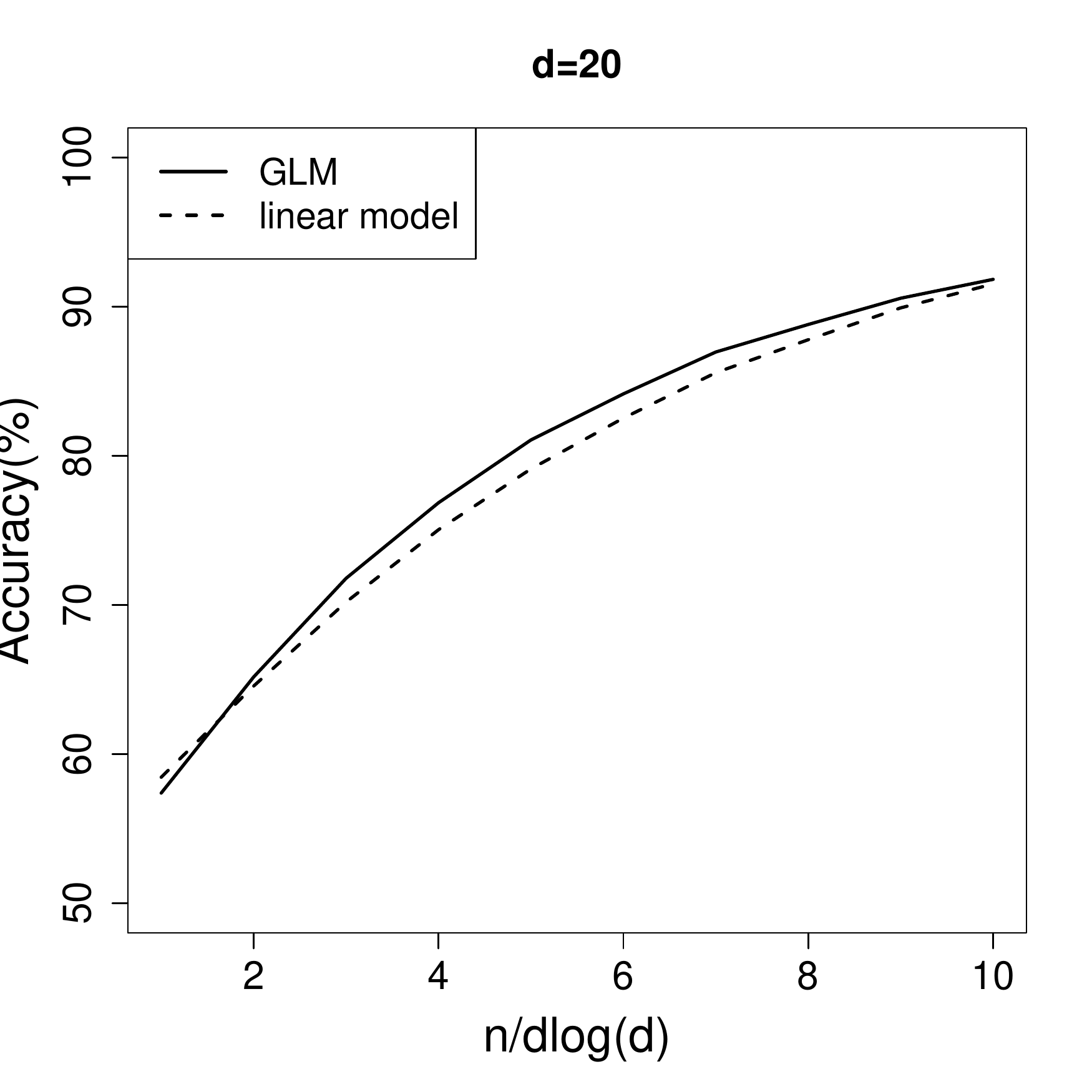}
\includegraphics[scale=0.4]{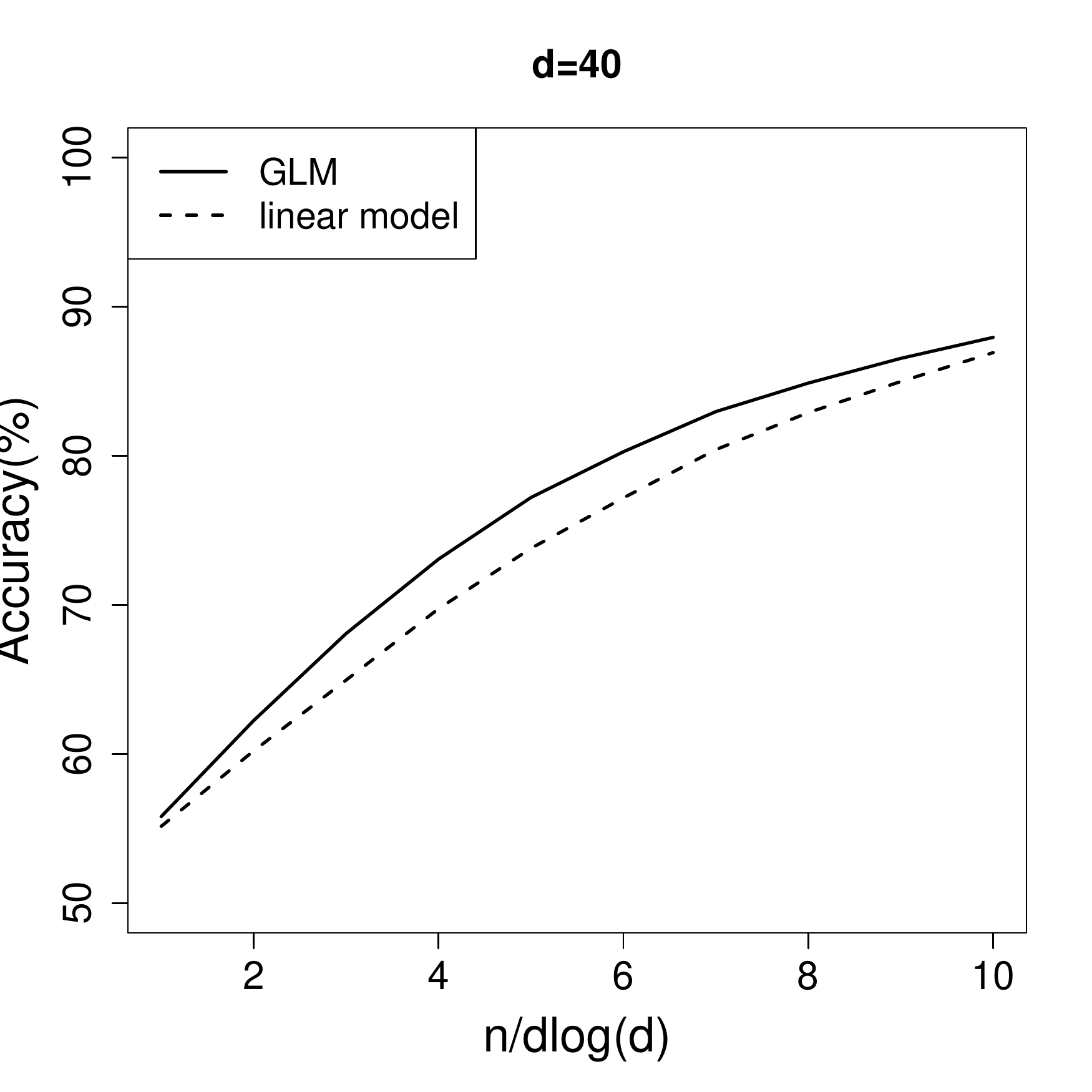}
\includegraphics[scale=0.4]{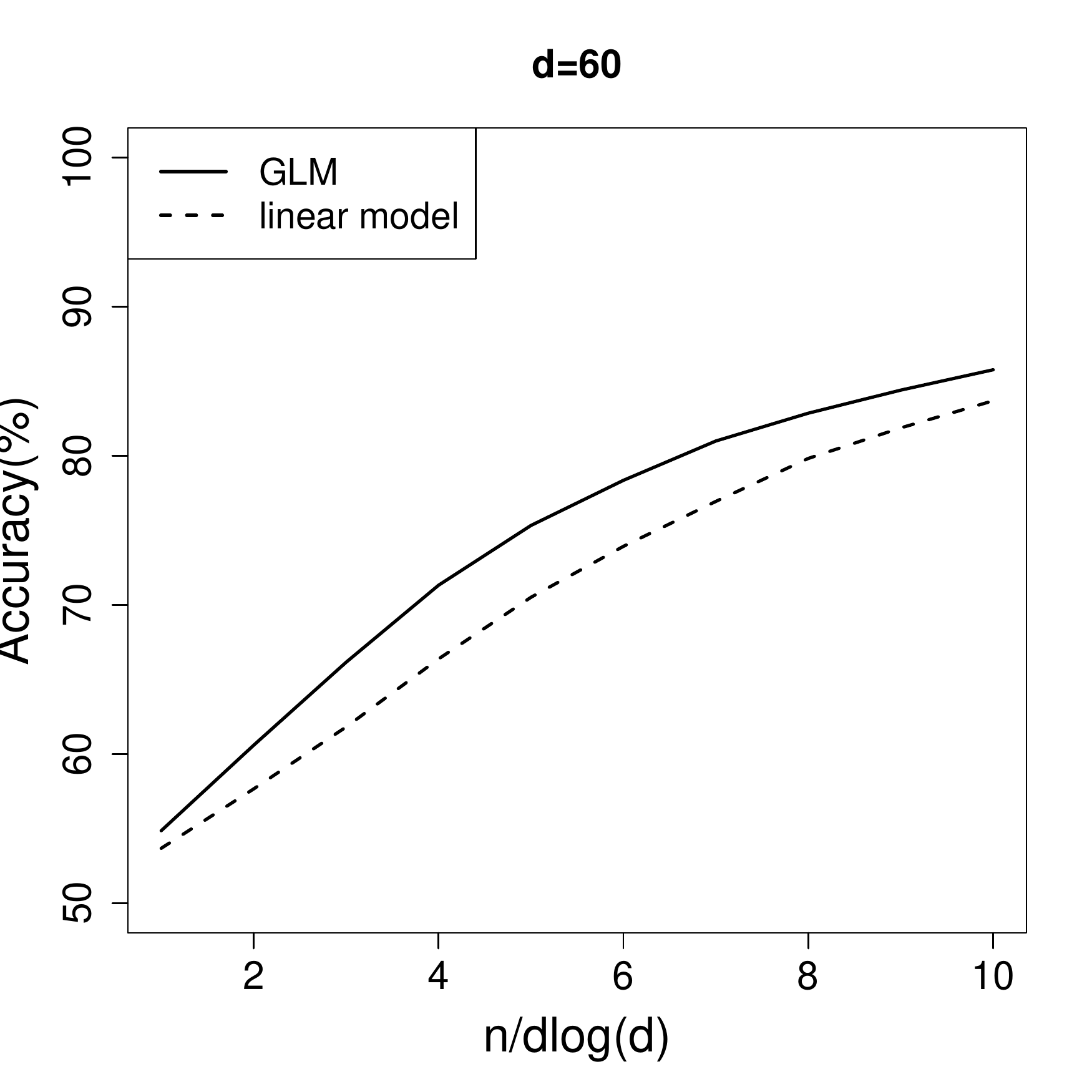}
\includegraphics[scale=0.4]{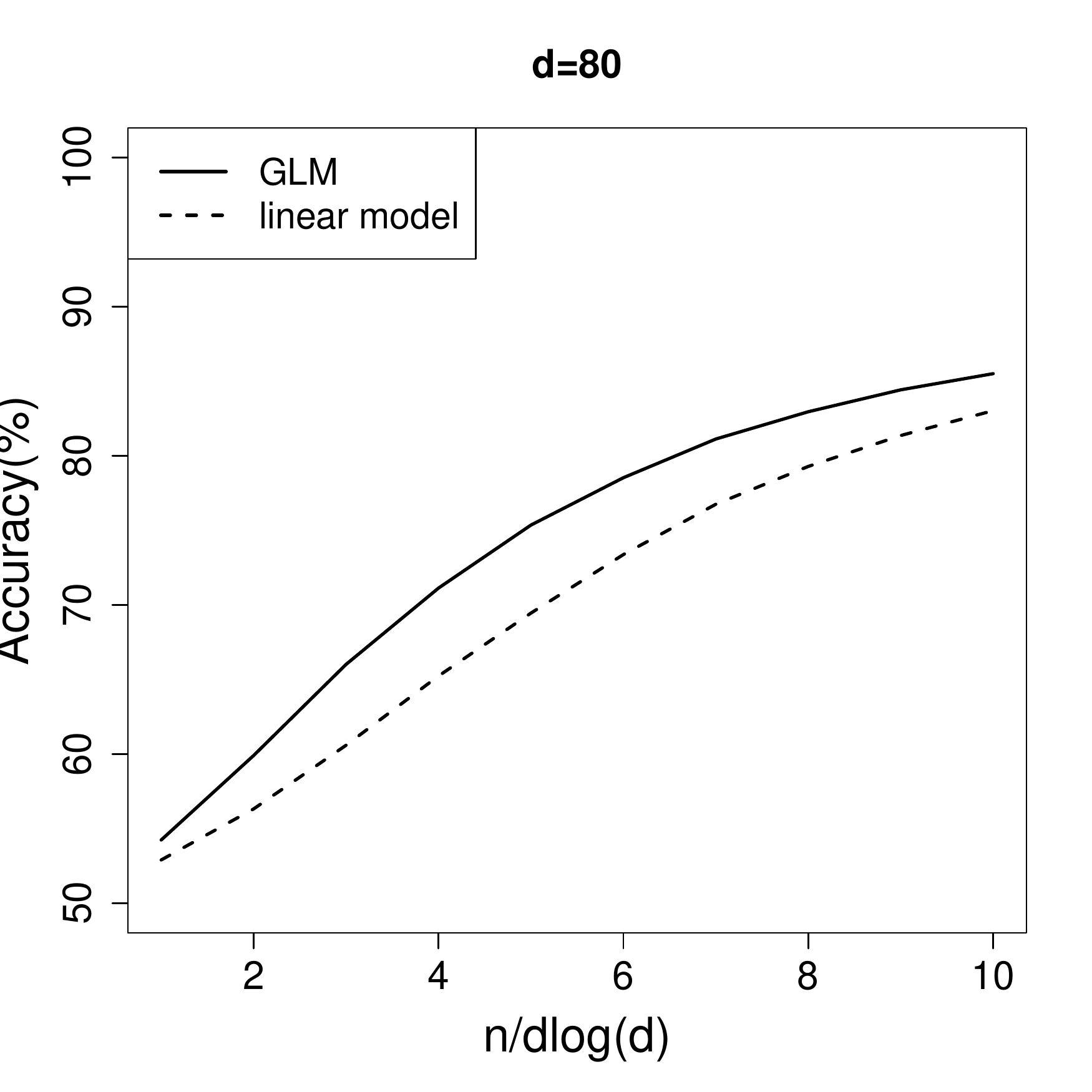}
\caption{Prediction accuracy  $1-\fnorm{\widehat \bT -\bT}^2/d^2$ in matrix completion for various dimension $d$.}
\label{fig:xx}
\end{figure}

We predict the ratings in two different ways. We first estimate the underlying $\bTheta^*$ with nuclear norm regularized logistic regression model. We set $\lambda=0.2\sqrt{d\log{d}/n}$ and derive the estimator $\hat{\bTheta}^{(1)}$ according to (\ref{eq:mincompletion}). We  estimate the rating matrix $\bT$ by $\hat \bT^{(1)}$ as defined below:  \[\hat{\bT}^{(1)}_{ij}=\begin{cases}1\quad\text{if}\quad\hat{\bTheta}^{(1)}_{ij}\geq 0\\0\quad\text{else}\end{cases}.\]
The second method is to estimate $\bTheta^*$ with nuclear norm regularized linear model. Again, we take the tuning parameter $\lambda=0.2\sqrt{d\log{d}/n}$ and derive the estimator $\hat{\bTheta}^{(2)}$ as follows:
\begin{equation}
\hat{\bTheta}^{(2)}=\argmin\limits_{\norm{\bTheta}_{\infty}\leq R} \left\{\frac{1}{n}\sum_{i=1}^n \left(Y_i-\inn{\bTheta^*,\bX_i}\right)^2+\lambda\norm{\bTheta}_N\right\}.
\end{equation}
To estimate the rating matrix $\bT$, we use \[\hat{T}^{(2)}_{ij}=\begin{cases}1\quad\text{if}\quad \hat{\bTheta}^{(2)}_{ij}\geq 0.5\\0\quad\text{else}\end{cases}.\]
The experiment is repeated for 100 times. In each repetition, we record the prediction accuracy as $1-\fnorm{\widehat \bT^{(k)}-\bT}^2/d^2$ for $k=1$ and $2$, which is the proportion of correct predictions. We plot the average prediction accuracy in Figure \ref{fig:xx}.

We use solid lines to denote the prediction accuracy achieved by regularized GLM and we use dotted lines to denote the accuracy achieved by regularized linear model. We can see from Figure \ref{fig:xx} that no matter how the dimension changes, the solid lines are always above the dotted lines, showing that the generalized model always outperforms the linear model with categorical response. This validates our motivation to use the generalized model in matrix recovery problems with categorical outcomes.

\section{Real Data Analysis}\label{sec:4}

In this section, we apply generalized trace regression with nuclear norm regularization to stock return prediction and image classification. The former can be regarded as a reduced rank regression and the latter can be seen as the categorical responses with matrix inputs.  The results demonstrate the advantage of recruiting nuclear norm penalty compared with no penalty or using $\ell_1$-norm regularization.

\subsection{Stock Return Prediction}

In this subsection we aim to predict the sign of the one-day forward stock return, i.e., whether the price of the stock will rise or fall in the next day. We pick $19$ individual stocks as our objects of study: \texttt{AAPL, BAC, BRK-B, C, COP, CVX, DIS, GE, GOOGL, GS, HON, JNJ, JPM, MRK, PFE, UNH, V, WFC} and \texttt{XOM}. These are the largest holdings of Vanguard ETF in technology, health care, finance, energy, industrials and consumer. We also include $\texttt{S\&P500}$ in our pool of stocks since it represents the market portfolio and should help the prediction. Therefore, we have $d_1 = 20$ stocks in total. We collect the daily returns of these stocks from 01/01/13 to 8/31/2017 and divide them into the training set (2013-2014), the evaluation set (2015) and the testing set (2016-2017). The sample sizes of the training, evaluation and testing sets are $n_1=504, n_2= 252$ and $n_3 = 420$ respectively.

 We fit a generalized reduced-rank regression model \eqref{eq:2.8} based on the moving average (MA) of returns of each stock in the past 1 day, 3 days, 5 days, 10 days and 20 days. Hence, the dimension of $\bx_i$ is $20 \times 5 = 100$. Let $\by_i\in\mathbb{R}^{20}$ be the sign of returns of the selected stocks on the $(i+1)$th day. We assume that $\bTheta^*\in\mathbb{R}^{20\times100}$ is a near low-rank matrix, considering high correlations across the returns of the selected stocks.  We tune $\lambda$ for the best performance on the evaluation data. When we predict on the test set, we will update $\widehat\bTheta$ on a monthly basis, i.e., for each month in the testing set, we refit \eqref{eq:2.8} based on the data in the most recent three years. Given an estimator $\hat{\bTheta}$, our prediction $\widehat\by_j$ are the signs of $(\widehat\bTheta^T\bx_j)$.

We have two baseline models in our analysis. The first one is the deterministic bet (DB): if a stock has more positive returns than negative ones in the training set, we always predict positive returns; otherwise, we always predict negative returns. The second one is the generalized RRR without any nuclear norm regularization. We use this baseline to demonstrate the advantage of incorporating nuclear norm regularization.

\begin{table}
\newcommand{\tabincell}[2]{\begin{tabular}{@{}#1@{}}#2\end{tabular}}
\centering
{
\begin{tabular}{|c|c|c|c|}
\hline
Stock & DB & \tabincell{c}{Prediction Accuracy \\with Regularization} & \tabincell{c}{Prediction Accuracy \\without Regularization} \\ \hline
AAPL & 55.13 & 51.07 & 51.07 \\
BAC & 47.26 & 49.88 & 49.64 \\
BRK-B & 54.18 & 59.90 & 59.90 \\
C & 52.98 & 51.55 & 51.07 \\
COP & 47.49 & 54.18 & 54.18 \\
CVX & 48.69& 55.37 & 54.18 \\
DIS &  49.40& 56.80 & 56.80 \\
GE & 48.45& 55.61 & 56.09 \\
GOOGL & 53.94 & 52.74 & 52.74\\
GS &  52.74 & 53.22 & 47.49 \\
HON &  56.09 & 51.55 & 51.31\\
JNJ &   51.79 & 54.65 & 53.70\\
JPM &   52.27 & 53.94 & 47.02\\
MRK &  51.55 & 51.31 & 51.31\\
PFE &  49.40 & 52.27 & 49.40\\
UNH & 52.74  & 53.70 & 52.74\\
V &   56.09 & 58.00 & 58.23\\
WFC &  49.16 & 52.74 & 50.12\\
XOM &  48.21 & 54.42 & 53.46\\
SPY & 54.89 & 54.89 & 54.42 \\\hline
Average & 51.62 & 53.89 & 52.74 \\\hline
\end{tabular}}
\caption{Prediction Result of 20 selected stocks.(Unit: \%)}
\label{tab:finance}
\end{table}

From Table \ref{tab:finance}, we can see that the nuclear norm penalized model yields an average accuracy of 53.89\% while the accuracy of the unpenalized model and DB are 52.74\% and 51.62\%. Note that the penalized model performs the same as or better than the unpenalized model in 18 out of 20 stocks. When compared with the DB, the penalized model performs better in 15 out of the 20 stocks. The improvement in the overall performance illustrates the advantage of using generalized RRR with nuclear norm regularization.

\subsection{CIFAR10 Dataset}

Besides the application in finance, we also apply our model to the well-known CIFAR10 dataset in image classification. The CIFAR10 dataset has 60,000 colored $32\times 32$ images in 10 classes: the airplane, automobile, bird, cat, dog, deer, dog, frog, horse, ship and truck. There are 3 channels (red, green and blue) in each figure, hence each image is stored as a $32\times96$ matrix. We represent the 10 classes with the numbers 0,1, \ldots, 9. The training data contains 50,000 figures and the testing data contains 10,000 figures. In our work, we only use 10,000 samples to train the model.

We construct and train a convolutional neural networks (CNN) with $\ell_1$ norm and nuclear norm regularizations on $\bTheta$ respectively to learn the pattern of the figures. The structure of the CNN follows the online tutorial from TensorFlow\footnote{The code can be downloaded from \url{https://github.com/tensorflow/models/tree/master/tutorials/image/cifar10}. The tutorial can be found at \url{https://www.tensorflow.org/tutorials/deep_cnn}}. It extracts a $384$-dimensional feature vector from each image and maps it to 10 categories through logistic regression with a $384\times 10$ coefficient matrix. Here to exploit potential matrix structure of the features, we reshape this 384-dimensional feature vector into a $24\times 16$ matrix and map it to one of the ten categories through generalized trace regression with ten $24 \times 16$ coefficient matrices. We impose nuclear norm and $\ell_1$-norm regularizations on $\bTheta$ on coefficient matrices respectively and we summarize our results in Table \ref{tab:4} below.

\begin{table}[htbp]
\centering
\begin{tabular}{|c|c|c|c|c|c|c|}
\hline
$\lambda$ & 0 & 0.02 & 0.05 & 0.1 & 0.2 & 0.3 \\ \hline
nuclear penalty & 74.30\% & 76.04\% & 76.17\% & 75.29\% & 74.45\% & 73.46\% \\ \hline
$\lambda$ & 0 & 0.001 & 0.002 & 0.005 & 0.008 & 0.01  \\ \hline
$\ell_1$ penalty & 74.30\% & 75.70\% & 75.90\% & 75.53\% & 75.37\% & 75.22\%  \\ \hline
\end{tabular}
\caption{Prediction accuracy in CIFAR10 under different $\lambda$ with different penalties with convolutional neural network.}
\label{tab:4}
\end{table}

The results show that both regularization methods promote the prediction accuracy while nuclear norm regularization again outperforms $\ell_1$ norm. The main reason might be that there is low-rankness instead of sparsity lying in the deep features extracted by neural network.

\bibliographystyle{plainnat}
\bibliography{ref}

\begin{thebibliography}{41}
\providecommand{\natexlab}[1]{#1}
\providecommand{\url}[1]{\texttt{#1}}
\expandafter\ifx\csname urlstyle\endcsname\relax
  \providecommand{\doi}[1]{doi: #1}\else
  \providecommand{\doi}{doi: \begingroup \urlstyle{rm}\Url}\fi

\bibitem[Ahn and Reinsel(1994)]{ARe94}
Sung~K Ahn and Gregory~C Reinsel.
\newblock Estimation of partially nonstationary vector autoregressive models
  with seasonal behavior.
\newblock \emph{Journal of Econometrics}, 62\penalty0 (2):\penalty0 317--350,
  1994.

\bibitem[Anderson(1951)]{And51}
Theodore~Wilbur Anderson.
\newblock Estimating linear restrictions on regression coefficients for
  multivariate normal distributions.
\newblock \emph{The Annals of Mathematical Statistics}, pages 327--351, 1951.

\bibitem[Belloni et~al.(2012)Belloni, Chen, Chernozhukov, and Hansen]{BCC12}
Alexandre Belloni, Daniel Chen, Victor Chernozhukov, and Christian Hansen.
\newblock Sparse models and methods for optimal instruments with an application
  to eminent domain.
\newblock \emph{Econometrica}, 80\penalty0 (6):\penalty0 2369--2429, 2012.

\bibitem[Belloni et~al.(2017)Belloni, Chernozhukov, Fern{\'a}ndez-Val, and
  Hansen]{belloni2017program}
Alexandre Belloni, Victor Chernozhukov, Ivan Fern{\'a}ndez-Val, and Christian
  Hansen.
\newblock Program evaluation and causal inference with high-dimensional data.
\newblock \emph{Econometrica}, 85\penalty0 (1):\penalty0 233--298, 2017.

\bibitem[Bhaskar and Javanmard(2015)]{bhaskar20151}
Sonia~A Bhaskar and Adel Javanmard.
\newblock 1-bit matrix completion under exact low-rank constraint.
\newblock In \emph{Information Sciences and Systems (CISS), 2015 49th Annual
  Conference on}, pages 1--6. IEEE, 2015.

\bibitem[Cai and Zhou(2013)]{cai2013max}
Tony Cai and Wen-Xin Zhou.
\newblock A max-norm constrained minimization approach to 1-bit matrix
  completion.
\newblock \emph{Journal of Machine Learning Research}, 14\penalty0
  (1):\penalty0 3619--3647, 2013.

\bibitem[Candes and Tao(2007)]{candes2007dantzig}
Emmanuel Candes and Terence Tao.
\newblock The dantzig selector: Statistical estimation when p is much larger
  than n.
\newblock \emph{The Annals of Statistics}, pages 2313--2351, 2007.

\bibitem[Caner and Fan(2015)]{caner2015hybrid}
Mehmet Caner and Qingliang Fan.
\newblock Hybrid generalized empirical likelihood estimators: Instrument
  selection with adaptive lasso.
\newblock \emph{Journal of Econometrics}, 187\penalty0 (1):\penalty0 256--274,
  2015.

\bibitem[Chan et~al.(2015)Chan, Yau, and Zhang]{CYZ15}
Ngai~Hang Chan, Chun~Yip Yau, and Rong-Mao Zhang.
\newblock Lasso estimation of threshold autoregressive models.
\newblock \emph{Journal of Econometrics}, 189\penalty0 (2):\penalty0 285--296,
  2015.

\bibitem[Chen et~al.(2013)Chen, Dong, and Chan]{CDC13}
Kun Chen, Hongbo Dong, and Kung-Sik Chan.
\newblock Reduced rank regression via adaptive nuclear norm penalization.
\newblock \emph{Biometrika}, 100\penalty0 (4):\penalty0 901--920, 2013.

\bibitem[Chen et~al.(2001)Chen, Donoho, and Saunders]{chen2001atomic}
Scott~Shaobing Chen, David~L Donoho, and Michael~A Saunders.
\newblock Atomic decomposition by basis pursuit.
\newblock \emph{SIAM review}, 43\penalty0 (1):\penalty0 129--159, 2001.

\bibitem[Davenport et~al.(2014)Davenport, Plan, van~den Berg, and
  Wootters]{davenport20141}
Mark~A Davenport, Yaniv Plan, Ewout van~den Berg, and Mary Wootters.
\newblock 1-bit matrix completion.
\newblock \emph{Information and Inference}, 3\penalty0 (3):\penalty0 189--223,
  2014.

\bibitem[{Fan} et~al.(2016){Fan}, {Wang}, and {Zhu}]{fan2016shrinkage}
J.~{Fan}, W.~{Wang}, and Z.~{Zhu}.
\newblock {A Shrinkage Principle for Heavy-Tailed Data: High-Dimensional Robust
  Low-Rank Matrix Recovery}.
\newblock \emph{ArXiv e-prints}, March 2016.

\bibitem[Fan and Li(2001)]{fan2001variable}
Jianqing Fan and Runze Li.
\newblock Variable selection via nonconcave penalized likelihood and its oracle
  properties.
\newblock \emph{Journal of the American statistical Association}, 96\penalty0
  (456):\penalty0 1348--1360, 2001.

\bibitem[Fan and Lv(2008)]{fan2008sure}
Jianqing Fan and Jinchi Lv.
\newblock Sure independence screening for ultrahigh dimensional feature space.
\newblock \emph{Journal of the Royal Statistical Society: Series B (Statistical
  Methodology)}, 70\penalty0 (5):\penalty0 849--911, 2008.

\bibitem[Fan and Lv(2011)]{fan2011nonconcave}
Jianqing Fan and Jinchi Lv.
\newblock Nonconcave penalized likelihood with np-dimensionality.
\newblock \emph{IEEE Transactions on Information Theory}, 57\penalty0
  (8):\penalty0 5467--5484, 2011.

\bibitem[Fan et~al.(2015)Fan, Liu, Sun, and Zhang]{FLS15}
Jianqing Fan, Han Liu, Qiang Sun, and Tong Zhang.
\newblock Tac for sparse learning: Simultaneous control of algorithmic
  complexity and statistical error.
\newblock \emph{arXiv preprint arXiv:1507.01037}, 2015.

\bibitem[Fang et~al.(2015)Fang, Liu, Toh, and Zhou]{fang2015max}
Ethan~X Fang, Han Liu, Kim-Chuan Toh, and Wen-Xin Zhou.
\newblock Max-norm optimization for robust matrix recovery.
\newblock \emph{Mathematical Programming}, pages 1--31, 2015.

\bibitem[Geweke(1996)]{Gew96}
John Geweke.
\newblock Bayesian reduced rank regression in econometrics.
\newblock \emph{Journal of econometrics}, 75\penalty0 (1):\penalty0 121--146,
  1996.

\bibitem[Gupta et~al.(2010)Gupta, Nowak, and Recht]{gupta2010sample}
Ankit Gupta, Robert Nowak, and Benjamin Recht.
\newblock Sample complexity for 1-bit compressed sensing and sparse
  classification.
\newblock In \emph{Information Theory Proceedings (ISIT), 2010 IEEE
  International Symposium on}, pages 1553--1557. IEEE, 2010.

\bibitem[Hansen and Kozbur(2014)]{hansen2014instrumental}
Christian Hansen and Damian Kozbur.
\newblock Instrumental variables estimation with many weak instruments using
  regularized jive.
\newblock \emph{Journal of Econometrics}, 182\penalty0 (2):\penalty0 290--308,
  2014.

\bibitem[Izenman(1975{\natexlab{a}})]{Ala75}
Alan~Julian Izenman.
\newblock Reduced-rank regression for the multivariate linear model.
\newblock \emph{Journal of multivariate analysis}, 5\penalty0 (2):\penalty0
  248--264, 1975{\natexlab{a}}.

\bibitem[Izenman(1975{\natexlab{b}})]{Ize75}
Alan~Julian Izenman.
\newblock Reduced-rank regression for the multivariate linear model.
\newblock \emph{Journal of multivariate analysis}, 5\penalty0 (2):\penalty0
  248--264, 1975{\natexlab{b}}.

\bibitem[Kleibergen and Paap(2006)]{KPa06}
Frank Kleibergen and Richard Paap.
\newblock Generalized reduced rank tests using the singular value
  decomposition.
\newblock \emph{Journal of econometrics}, 133\penalty0 (1):\penalty0 97--126,
  2006.

\bibitem[Kock and Callot(2015)]{kock2015oracle}
Anders~Bredahl Kock and Laurent Callot.
\newblock Oracle inequalities for high dimensional vector autoregressions.
\newblock \emph{Journal of Econometrics}, 186\penalty0 (2):\penalty0 325--344,
  2015.

\bibitem[Koltchinskii et~al.(2011)Koltchinskii, Lounici, and
  Tsybakov]{koltchinskii2011nuclear}
Vladimir Koltchinskii, Karim Lounici, and Alexandre~B Tsybakov.
\newblock Nuclear-norm penalization and optimal rates for noisy low-rank matrix
  completion.
\newblock \emph{The Annals of Statistics}, pages 2302--2329, 2011.

\bibitem[Lee et~al.(2014)Lee, Lou, Chen, Chen, Lin, Chiang, and Chen]{LLC14}
Eric~L Lee, Jing-Kai Lou, Wei-Ming Chen, Yen-Chi Chen, Shou-De Lin, Yen-Sheng
  Chiang, and Kuan-Ta Chen.
\newblock Fairness-aware loan recommendation for microfinance services.
\newblock In \emph{Proceedings of the 2014 International Conference on Social
  Computing}, page~3. ACM, 2014.

\bibitem[Ludvigson and Ng(2009)]{ludvigson2009macro}
Sydney~C Ludvigson and Serena Ng.
\newblock Macro factors in bond risk premia.
\newblock \emph{The Review of Financial Studies}, 22\penalty0 (12):\penalty0
  5027--5067, 2009.

\bibitem[Marshak(1950)]{Mar50}
Jacob Marshak.
\newblock \emph{Statistical inference in economics: an introduction}.
\newblock John Wiley $\&$ Sons, 1950.

\bibitem[Negahban and Wainwright(2011)]{negahban2011estimation}
Sahand Negahban and Martin~J Wainwright.
\newblock Estimation of (near) low-rank matrices with noise and
  high-dimensional scaling.
\newblock \emph{The Annals of Statistics}, pages 1069--1097, 2011.

\bibitem[Negahban and Wainwright(2012)]{NWa12}
Sahand Negahban and Martin~J Wainwright.
\newblock Restricted strong convexity and weighted matrix completion: Optimal
  bounds with noise.
\newblock \emph{Journal of Machine Learning Research}, 13\penalty0
  (1):\penalty0 1665--1697, 2012.

\bibitem[Negahban et~al.(2011)Negahban, Ravikumar, Wainwright, and Yu]{NRW11}
Sahand Negahban, Pradeep Ravikumar, Martin~J Wainwright, and Bin Yu.
\newblock A unified framework for high-dimensional analysis of m-estimators
  with decomposable regularizers.
\newblock In \emph{Adv. Neural Inf. Proc. Sys.(NIPS)}. Citeseer, 2011.

\bibitem[Plan and Vershynin(2013{\natexlab{a}})]{plan2013one}
Yaniv Plan and Roman Vershynin.
\newblock One-bit compressed sensing by linear programming.
\newblock \emph{Communications on Pure and Applied Mathematics}, 66\penalty0
  (8):\penalty0 1275--1297, 2013{\natexlab{a}}.

\bibitem[Plan and Vershynin(2013{\natexlab{b}})]{plan2013robust}
Yaniv Plan and Roman Vershynin.
\newblock Robust 1-bit compressed sensing and sparse logistic regression: A
  convex programming approach.
\newblock \emph{IEEE Transactions on Information Theory}, 59\penalty0
  (1):\penalty0 482--494, 2013{\natexlab{b}}.

\bibitem[Raskutti et~al.(2010)Raskutti, Wainwright, and Yu]{RWY10}
Garvesh Raskutti, Martin~J Wainwright, and Bin Yu.
\newblock Restricted eigenvalue properties for correlated gaussian designs.
\newblock \emph{Journal of Machine Learning Research}, 11\penalty0
  (Aug):\penalty0 2241--2259, 2010.

\bibitem[Stock and Watson(2002)]{stock2002forecasting}
James~H Stock and Mark~W Watson.
\newblock Forecasting using principal components from a large number of
  predictors.
\newblock \emph{Journal of the American statistical association}, 97\penalty0
  (460):\penalty0 1167--1179, 2002.

\bibitem[Tibshirani(1996)]{tibshirani1996regression}
Robert Tibshirani.
\newblock Regression shrinkage and selection via the lasso.
\newblock \emph{Journal of the Royal Statistical Society. Series B
  (Methodological)}, pages 267--288, 1996.

\bibitem[Velu and Reinsel(2013)]{VRe13}
Raja Velu and Gregory~C Reinsel.
\newblock \emph{Multivariate reduced-rank regression: theory and applications},
  volume 136.
\newblock Springer Science \& Business Media, 2013.

\bibitem[Vershynin(2010)]{Ver10}
Roman Vershynin.
\newblock Introduction to the non-asymptotic analysis of random matrices.
\newblock \emph{arXiv preprint arXiv:1011.3027}, 2010.

\bibitem[Zhang et~al.(2010)]{zhang2010nearly}
Cun-Hui Zhang et~al.
\newblock Nearly unbiased variable selection under minimax concave penalty.
\newblock \emph{The Annals of statistics}, 38\penalty0 (2):\penalty0 894--942,
  2010.

\bibitem[Zou and Li(2008)]{zou2008one}
Hui Zou and Runze Li.
\newblock One-step sparse estimates in nonconcave penalized likelihood models.
\newblock \emph{Annals of statistics}, 36\penalty0 (4):\penalty0 1509, 2008.

\end{thebibliography}

\section{Proofs and Technical Lemmas}

\subsection{\bf Proof for Theorem \ref{thm:1}}

We follow the proof scheme of Lemma B.4 in \cite{FLS15}. We first construct a middle point $\hat{\bTheta}_{\eta}=\bTheta^*+\eta(\hat{\bTheta}-\bTheta^*)$ such that we choose $\eta = 1$ when $\fnorm{\hat{\bTheta} - \bTheta^*} \le \ell$ and $\eta = \ell / \fnorm{\hat{\bTheta} - \bTheta^*}$ when $\fnorm{\hat{\bTheta} - \bTheta^*} > \ell$. Here, $\ell$ will be determined later. We denote the Frobenius ball $\cN=\{\bTheta: \fnorm{\bTheta-\bTheta^*}\leq\ell\}$. For simplicity, we let $\hat{\bDelta}=\hat{\bTheta}-\bTheta^*$ and $\hat{\bDelta}_{\eta}=\hat{\bTheta}_{\eta}-\bTheta^*$ in the remainder of the proof.

According to Negahban et al. (2012), when $\lambda\geq 2\opnorm{n^{-1}\sum\limits_{i=1}^n \left[b'(\inn{\bX_i, \bTheta^*}) - Y_i\right]\cdot\bX_i}$, $\hat{\bDelta}$ falls in the following cone:
 \[\mathcal{C}(\mathcal{M},\overline{\mathcal{M}}^{\perp},\bTheta^*):=\Bigl\{\norm{\bDelta_{\overline{\mathcal{M}}^{\perp}}}_N\leq3\norm{\bDelta_{\overline{\mathcal{M}}}}_N+4\sum\limits_{j\geq r+1}\sigma_j(\bTheta^*)\Bigr\}.\]
Since $\hat{\bDelta}_{\eta}$ is parallel to $\hat{\bDelta}$, $\hat{\bDelta}_{\eta}$ also falls in this cone. Given $\nnorm{\hat{\bDelta}_{\eta}} \leq \ell$ and LRSC($\cC, \cN, \kappa_{\ell}, \tau_{\ell}$) of $\cL(\bTheta)$, we have
\begin{equation}
\kappa_{\ell}\fnorm{\hat{\bDelta}_{\eta}}^2 - \tau_{\ell}\leq\inn{\nabla\mathcal{L}(\hat{\bTheta}_{\eta}) -\nabla\mathcal{L}(\bTheta^*),\hat{\bDelta}_{\eta}}=:D^s_{\mathcal{L}}(\hat{\bTheta}_{\eta}, \bTheta^*),
\end{equation}
where $D_{\mathcal{L}}(\bTheta_1,\bTheta_2)=\mathcal{L}(\bTheta_1)-\mathcal{L}(\bTheta_2)-\inn{\nabla\mathcal{L}(\bTheta_2),\bTheta_1-\bTheta_2}$ and $D^s_{\mathcal{L}}(\bTheta_1,\bTheta_2)=D_{\mathcal{L}}(\bTheta_1,\bTheta_2)+D_{\mathcal{L}}(\bTheta_2,\bTheta_1)$. By Lemma F.4 in \cite{FLS15}, $D^s_{\mathcal{L}}(\hat{\bTheta}_{\eta},\bTheta^*)\leq\eta\cdot D^s_{\mathcal{L}}(\hat{\bTheta},\bTheta^*).$
We thus have
\begin{equation}\label{eq:234}
\kappa_{\ell}\fnorm{\hat{\bDelta}_{\eta}}^2 - \tau_{\ell}\leq D^s_{\mathcal{L}}(\hat{\bTheta}_{\eta}, \bTheta^*)\leq \eta D^s_{\mathcal{L}}(\hat{\bTheta}, \bTheta^*)=\inn{\nabla\mathcal{L}(\hat{\bTheta})-\nabla\mathcal{L}(\bTheta^*),\hat{\bDelta}_{\eta}}.
\end{equation}
Since $\hat{\bTheta}$ is the minimizer of the loss, we shall have the optimality condition $\nabla\mathcal{L}(\hat{\bTheta})+\lambda\bxi=\bzero$ for some subgradient $\bxi$ of the $\nnorm{{\bTheta}}$ at $\bTheta = \hat \bTheta$. Therefore, (\ref{eq:234}) simplifies to
\begin{equation}\label{eq:6.3}
\begin{split}
&\kappa_{\ell}\fnorm{\hat{\bDelta}_{\eta}} ^ 2 - \tau_{\ell} \leq-\inn{\nabla\mathcal{L}(\bTheta^*)+\lambda\bxi,\hat{\bDelta}_{\eta}}\leq 1.5\lambda\nnorm{\hat{\bDelta}_{\eta}} \leq  6\lambda\sqrt{2r}\norm{(\hat{\bDelta}_{\eta})_{\overline{\mathcal{M}}}}_F+6\lambda\sum\limits_{j\geq r+1} \sigma_j(\bTheta^*) \\
& \leq 6\lambda\sqrt{2r}\norm{\hat{\bDelta}_{\eta}}_F+6\lambda\sum\limits_{j\geq r+1} \sigma_j(\bTheta^*).
\end{split}
\end{equation}
For a threshold $\tau>0$, we choose $r=\# \{j \in \{1, 2, \ldots , d\} | \sigma_j(\bTheta^*) \ge \tau\}$. Then it follows that
\beq
	\label{eq:6.4}
	\sum\limits_{j \ge r+1} \sigma_j(\bTheta^*)\le \tau \sum\limits_{j \ge r+1} \frac{\sigma_j(\bTheta^*)}{\tau} \le \tau \sum\limits_{j \ge r+1} \bigl(\frac{\sigma_j(\bTheta^*)}{\tau}\bigr)^q \le \tau^{1-q}\sum\limits_{j \ge r+1}\sigma_j(\bTheta^*)^q \le \tau^{1-q}\rho.
\eeq
On the other hand, $\rho \ge \sum\limits_{j \le r} \sigma_j(\bTheta^*)^q \ge r\tau^q$, so $r \le \rho\tau^{-q}$. Choose $\tau = \lambda / \kappa_{\ell}$. Given \eqref{eq:6.3}, \eqref{eq:6.4} and $\tau_{\ell} = C_0 \rho \lambda^{2-q} / \kappa^{1-q}_{\ell}$ yields that for some constant $C_1$, $\fnorm{\widehat\bDelta_{\eta}} \le C_1\sqrt{\rho}(\lambda / \kappa_{\ell})^{1 - q / 2}$. If we choose $\ell > C_1\sqrt{\rho}(\lambda / \kappa_{\ell})^{1 - q / 2}$ in advance, we have $\bDelta_{\eta} = \bDelta$. Note that $\rank(\widehat\bDelta_{\overline\cM})\le 2r$; we thus have
\beq
	\label{eq:6.5}
	\begin{aligned}
		\nnorm{\widehat\bDelta} & \le \nnorm{(\widehat\bDelta)_{\overline\cM}}+\nnorm{(\widehat\bDelta)_{\overline\cM^{\perp}}}\le 4\nnorm{(\widehat\bDelta)_{\overline\cM}}+4\sum_{j\ge r+1}\sigma_j(\bTheta^*) \\
		& \le 4\sqrt{2r}\fnorm{\widehat\bDelta}+4\sum_{j\ge r+1}\sigma_j(\bTheta^*) \le 4\sqrt{\rho}\tau^{-\frac{q}{2}}\fnorm{\bDelta} + 4\rho \Bigl(\frac{\lambda}{\kappa_{\ell}} \Bigr)^{1-q} \le (4C_1 + 4)\rho \Bigl( \frac{\lambda}{\kappa_{\ell}}\Bigr)^{1 - q}.
	\end{aligned}
\eeq

\subsection{\bf Proof for Lemma \ref{lem:1}}

Let $\eta_i^*=\inn{\bTheta^*,\bX_i}$ and $\eta^*=\inn{\bTheta^*,\bX}$. Since $\mathbb{E}[(b'(\eta^*)-Y)\bX]=\mathbb{E}[b'(\eta^*)-Y]\cdot\mathbb{E}[\bX]=0$ due to independency, we have
\begin{equation}
\norm{\frac{1}{n}\sum_{i=1}^n(b'(\eta_i^*)-Y_i)\bX_i}_{\text{op}} 
=\norm{\frac{1}{n}\sum_{i=1}^n(b'(\eta_i^*)-Y_i)\bX_i-\mathbb{E}[(b'(\eta^*)-Y)\bX]}_{\text{op}} \\
\end{equation}
We use the covering argument to bound the above operator norm.

Let $\cS^d=\left\{\bu\in\mathbb{R}^d:\norm{\bu}_2=1\right\}$, $\cN^d$ be the $1/4$ covering on $\cS^d$ and $\Phi(\bA)=\sup\limits_{\substack{\bu\in\cN^d \\ \bv\in\cN^d}}\bu^T\bA\bv$ for $\forall \bA\in\mathbb{R}^{d\times d}$.

We claim that
\begin{equation}\label{eq:1}
\norm{\bA}_{\text{op}}\leq\frac{16}{7}\Phi(\bA).
\end{equation}

To establish the above inequality, we shall notice that since $\mathcal{N}^{d}$ is a $1/4$
covering, for any given $\bu\in\cS^d,\bv\in\cS^d$, there is a $\tilde{\bu}\in\cN^d$ and $\tilde{\bv}\in\cN^d$ such that $\norm{\bu-\tilde{\bu}}\leq 1/4$ and $\norm{\bv-\tilde{\bv}}\leq 1/4$. Therefore,
\begin{align*}
\bu^T\bA\bv=&\tilde{\bu}^T\bA\tilde{\bv}+\tilde{\bu}^T\bA(\bv-\tilde{\bv})+(\bu-\tilde{\bu})^T\bA\tilde{\bv}+(\bu-\tilde{\bu})\bA(\bv-\tilde{\bv}) \\
\leq & \Phi(\bA)+\frac{1}{4}\norm{\bA}_{\text{op}}+\frac{1}{4}\norm{\bA}_{\text{op}}+\frac{1}{16}\norm{\bA}_{\text{op}} \\
=&\Phi(\bA)+\frac{9}{16}\norm{\bA}_{\text{op}}
\end{align*}
Take the supremum over all possible $\bu\in\cS^d,\bv\in\cS^d$, we have
\[\norm{\bA}_{\text{op}}=\sup\limits_{\substack{\bu\in\cS^d \\ \bv\in\cS^d}}\bu^T\bA\bv\leq\Phi(\bA)+\frac{9}{16}\norm{\bA}_{\text{op}} \]
and this leads to (\ref{eq:1}).

In the remaining of this proof, for fixed $\bu\in\cN^d$ and $\bv\in\cN^d$, denote $\bu^T\bX_i\bv$ by $Z_i$ and $\bu^TX\bv$ by $Z$ for convenience. According to the definition of sub-gaussian norm and sub-exponential norm, given the independence between the two terms, we have $\norm{\left[b'(\eta_i^*)-Y_i\right]Zi}_{\psi_1}\leq\norm{b'(\eta_i^*)-Y_i}_{\psi_2}\norm{Z_i}_{\psi_2}\leq \phi M\kappa_0$. Here, the reason why $\norm{b'(\eta_i^*)-Y_i}_{\psi_2}\leq\phi M$ is shown in the proof of Lemma \ref{lem:3}. By Proposition 5.16 (Bernstein-type inequality) in Vershynin (2010), it follows that for sufficiently small $t$,
\begin{equation}
\mathbb{P}\left(\left|\frac{1}{n}\sum_{i=1}^n (b'(\eta_i^*)-Y_i)Z_i-\mathbb{E}[(b'(\eta^*)-Y_i)Z]\right|>t\right)\leq2\exp{\left(-\frac{c_1nt^2}{\phi^2 M^2\kappa_0^2}\right)}
\end{equation}
where $c_1$ is a positive constant. 

Then the combination of the union bound over all points on $\cN^d\times\cN^d$ and (\ref{eq:1}) delivers
\begin{equation}\label{eq:2}
\mathbb{P}\left(\norm{\frac{1}{n}\sum_{i=1}^n (b'(\eta_i^*)-Y_i)Z_i-\mathbb{E}[(b'(\eta^*)-Y)Z]}_{\text{op}}>\frac{16}{7}t\right)\leq2\exp{\left(d\log{4}-\frac{c_1nt^2}{M^2\kappa_0^2}\right)}.
\end{equation}

In conclusion, if we choose $t\asymp\sqrt{d/n}$, we can find a constant $\gamma>0$ such that as long as $d/n<\gamma$, it holds that
\begin{equation}
P\left(\norm{\frac{1}{n}\sum_{i=1}^n(b'(\eta_i)-Y_i)\bX_i}_{\text{op}}>\nu\sqrt{\frac{d}{n}}\right)\leq c_1\cdot e^{-c_2d}.
\end{equation}
where $c_1$ and $c_2$ are constants.

\subsection{Proof for Lemma \ref{lem:2}}

In this proof, we will first show the RSC of $ \cL_n(\bTheta)$ at $\bTheta = \bTheta^*$ over the cone
\[
	\mathcal{C}(\mathcal{M}_r,\overline{\mathcal{M}}_r^{\perp},\bTheta^*)=\Bigl\{\bDelta\in\mathbb{R}^{d\times d}: \norm{\bDelta_{\overline{\mathcal{M}}_r^{\perp}}}_N\leq3\norm{\bDelta_{\overline{\mathcal{M}}_r}}_N+4\sum\limits_{j\geq r+1}\sigma_j(\bTheta^*)\Bigr\}
\]
for some $1\leq r\leq d$. Then, we will prove the LRSC of $ \cL_n(\bTheta)$ in a Frobenius norm neighborhood of $\bTheta^*$ with respect to the same cone.

\begin{enumerate}

\item An important inequality that leads to RSC of $ \cL_n(\bTheta)$ at $\bTheta = \bTheta^*$.

We first prove that the following inequality holds for all $\bDelta\in\mathbb{R}^{d\times d}$ with probability greater than $1-\exp(-c_1d)$:
\begin{equation}\label{eq:6.30}
\text{vec}(\bDelta)^T\cdot\hat{\bH}(\bTheta^*)\cdot\text{vec}(\bDelta)\geq\kappa\cdot\norm{\bDelta}_F^2-C_0\sqrt{\frac{d}{n}}\norm{\bDelta}_N^2.
\end{equation}

Let $\bDelta=\bU\bD\bV^T$ be the SVD of $\bDelta$. Then $\norm{\text{vec}(\bD)}_2=\norm{\bDelta}_F$ and $\norm{\text{vec}(\bD)}_1=\norm{\bDelta}_N$. It follows that
\begin{equation}\label{eq:6.31}
\begin{split}
&\text{vec}(\bDelta)^T\cdot \hat{\bH}(\bTheta^*)\cdot \text{vec}(\bDelta) \\
=&\frac{1}{n}\sum\limits_{i=1}^n \text{vec}(\bDelta)^T\cdot b''(\langle\bTheta^*, \bX_i\rangle)\cdot\text{vec}(\bX_i)\cdot\text{vec}(\bX_i)^T\cdot\text{vec}(\bDelta) \\
=&\frac{1}{n}\sum\limits_{i=1}^n \text{tr}(\sqrt{b''(\langle\bTheta^*, \bX_i\rangle)}\bX_i^T\bDelta)^2 =\frac{1}{n}\sum\limits_{i=1}^n \text{tr}(\sqrt{b''(\langle\bTheta^*, \bX_i\rangle)}\bX_i^T\bU\bD\bV^T)^2 \\
=&\frac{1}{n}\sum\limits_{i=1}^n \text{tr}(\sqrt{b''(\langle\bTheta^*, \bX_i\rangle)}\bV^T\bX_i^T\bU\bD)^2 =\frac{1}{n}\sum\limits_{i=1}^n \text{tr}(\tilde{\bX}_i^T\bD)^2=\text{vec}(\bD)^T\cdot\hat{\bSigma}_{\tilde{\bX}\tilde{\bX}}\cdot\text{vec}(\bD) \\
=&\text{vec}(\bD)^T\cdot\bSigma_{\tilde{\bX}\tilde{\bX}}\cdot\text{vec}(\bD)+\text{vec}(\bD)^T\cdot(\hat{\bSigma}_{\tilde{\bX}\tilde{\bX}}-\bSigma_{\tilde{\bX}\tilde{\bX}})\cdot\text{vec}(\bD) \\
\end{split}
\end{equation}

Here, $\tilde{\bX}_i=\sqrt{b''(\langle\bTheta^*, \bX_i\rangle)}\bU^T\bX_i\bV$, $\hat{\bSigma}_{\tilde{\bX}\tilde{\bX}}=n^{-1}\sum\limits_{i=1}^n\text{vec}(\tilde{\bX}_i)\cdot\text{vec}(\tilde{\bX}_i)^T$ and $\bSigma_{\tilde{\bX}\tilde{\bX}}=\mathbb{E}\hat{\bSigma}_{\tilde{\bX}\tilde{\bX}}$. 

To derive a lower bound for (\ref{eq:6.31}), we bound the first term from below and bound the second one from above.
\begin{equation}
\begin{split}
\lambda_{\text{min}}(\bSigma_{\tilde{\bX}\tilde{\bX}})=&\inf\limits_{\substack{\bW_1,\bW_2\in\mathbb{R}^{d\times d} \\ \norm{\bW_1}_F=\norm{\bW_2}_F=1}}\text{vec}(\bW_1)^T\cdot\bSigma_{\tilde{\bX}\tilde{\bX}}\cdot\text{vec}(\bW_2) \\
=&\inf\limits_{\substack{\bW_1,\bW_2\in\mathbb{R}^{d\times d} \\ \norm{\bW_1}_F=\norm{\bW_2}_F=1}}\mathbb{E}\left[\text{tr}(\sqrt{b''(\langle\bTheta^*, \bX_i\rangle)}\bW_1^T\bU^T\bX_i\bV)\cdot\text{tr}(\sqrt{b''(\langle\bTheta^*, \bX_i\rangle)}\bW_2^T\bU^T\bX_i\bV)\right] \\
=&\inf\limits_{\substack{\bW_1,\bW_2\in\mathbb{R}^{d\times d} \\ \norm{\bW_1}_F=\norm{\bW_2}_F=1}}\mathbb{E}\left[\text{tr}(\sqrt{b''(\langle\bTheta^*, \bX_i\rangle)}\bV\bW_1^T\bU^T\bX_i)\cdot\text{tr}(\sqrt{b''(\langle\bTheta^*, \bX_i\rangle)}\bV\bW_2^T\bU^T\bX_i)\right] \\
=&\inf\limits_{\substack{\bW_1,\bW_2\in\mathbb{R}^{d\times d} \\ \norm{\bW_1}_F=\norm{\bW_2}_F=1}}\text{vec}(\bU\bW_1\bV)\cdot \bH(\bTheta^*)\cdot \text{vec}(\bU\bW_2\bV) \\
=&\lambda_{\text{min}}(\bH(\bTheta^*))=\kappa
\end{split}
\end{equation}
Hence,
\begin{equation}\label{eq:5.14}
\text{vec}(\bDelta)^T\cdot \hat{\bSigma}_{\bX\bX}\cdot \text{vec}(\bDelta)\geq\kappa\norm{\bDelta}_F^2-\norm{\hat{\bSigma}_{\tilde{\bX}\tilde{\bX}}-\bSigma_{\tilde{\bX}\tilde{\bX}}}_{\infty}\norm{\bDelta}_N^2.
\end{equation}

Meanwhile, for some appropriate constants $c_3, c_4$ and $C_1$, we establish the following inequality, which serves as the key step to bound $\|\widehat\bSigma_{\tilde\bX\tilde\bX}-\bSigma_{\tilde\bX\tilde\bX}\|_{\infty}$.
		\beq
			\label{eq:6.33}
			\mathbb{P}\left(\Bigl | \sup_{\substack{\bu_1, \bu_2\in \cS^{d-1} \\ \bv_1, \bv_2 \in \cS^{d-1}} } \vec(\bu_1\bv_1^T)^T(\widehat\bSigma_{\tilde{\bX}\tilde{\bX}}- \bSigma_{\tilde{\bX}\tilde{\bX}})\vec(\bu_2\bv_2^T) \Bigr |> C_1\sqrt{\frac{d}{n}}\right) \le c_3\exp(-c_4d).
		\eeq
		We apply the covering argument to prove the claim above. Denote the $1/8-$net of $\cS^{d-1}$ by $\cN^{d-1}$. For any $\bA \in \RR^{d^2\times d^2}$, define
		\[
			\Phi(\bA):=\sup_{\substack{\bu_1, \bu_2\in \cS^{d-1} \\ \bv_1, \bv_2 \in \cS^{d-1}} } \vec(\bu_1\bv_1^T)^T\bA\vec(\bu_2\bv_2^T)
		\]
		and
		\[
			\Phi_{\cN}(\bA):=\sup_{\substack{\bu_1, \bu_2\in \cN^{d-1} \\ \bv_1, \bv_2 \in \cN^{d-1}} } \vec(\bu_1\bv_1^T)^T\bA\vec(\bu_2\bv_2^T).
		\]
		
		Note that for any $\bu_1,\bv_1, \bu_2,\bv_2 \in \cS^{d-1}$, there exist $\overline\bu_1, \overline\bv_1, \overline\bu_2, \overline\bv_2 \in \cN^{d-1}$  such that $\ltwonorm{\bu_i-\overline\bu_i}\le 1/8$ and $\ltwonorm{\bv_i-\overline\bv_i}\le 1/8$ for $i=1,2$. Then it follows that
		\beq
			\begin{aligned}
				& \vec (\bu_1\bv_1^T)^T\bA\vec(\bu_2\bv_2^T) \\
				& =\vec(\overline\bu_1\overline\bv_1^T)^T\bA\vec(\overline\bu_2\overline\bv_2^T)+\vec(\bu_1(\bv_1-\overline\bv_1)^T)^T\bA\vec(\overline\bu_2\overline\bv_2^T)+\vec((\bu_1-\overline\bu_1)\overline\bv_1^T)^T\bA\vec(\overline\bu_2\overline\bv_2^T)\\
				& +\vec(\bu_1\bv_1^T)^T\bA\vec(\bu_2(\bv_2-\overline\bv_2)^T)+ \vec(\bu_1\bv_1^T)^T\bA\vec((\bu_2-\overline\bu_2)\overline\bv_2^T)\\
				& +\vec((\bu_1-\overline\bu_1)\bv_1^T)^T\bA\vec((\bu_2-\overline\bu_2)\overline\bv_2^T)+\vec(\bu_1(\bv_1-\overline\bv_1)^T)^T\bA\vec((\bu_2-\overline\bu_2)\overline\bv_2^T)\\
				& +\vec((\bu_1-\overline\bu_1)\bv_1^T)^T\bA\vec(\bu_2(\bv_2-\overline\bv_2)^T)+\vec(\bu_1(\bv_1-\overline\bv_1)^T)^T\bA\vec(\bu_2(\bv_2-\overline\bv_2)^T) \\
				& \le \Phi_{\cN}(\bA)+\frac{1}{2}\Phi(\bA)+\frac{1}{16}\Phi(\bA).
			\end{aligned}
		\eeq
		So we have $\Phi(\bA) \le (16/7)\Phi_{\cN}(\bA)$. For any $\bu_1, \bu_2\in\cS^{d-1}$ and $\bv_1, \bv_2\in \cS^{d-1}$, we know from Lemma 5.14 in \cite{Ver10} that
		\begin{equation}
		\begin{split}
			\| & \inn{\bu_1\bv_1', \tilde{\bX}_i}  \inn{\bu_2\bv_2', \tilde{\bX}_i}\|_{\psi_1}\le \frac{1}{2}(\|\inn{\bu_1\bv_1', \tilde{\bX}_i}^2\|_{\psi_1}+\|\inn{\bu_2\bv_2', \tilde{\bX}_i}^2\|_{\psi_1}) \\
			& \le\norm{\inn{\bu_1\bv_1^T, \sqrt{b''(\inn{\bTheta^*, \bX_i})}\bU^T\bX_i\bV}}_{\psi_2}^2+\norm{\inn{\bu_2\bv_2^T, \sqrt{b''(\inn{\bTheta^*, \bX_i})}\bU^T\bX_i\bV}}_{\psi_2}^2 \leq  2M\kappa_0^2.
			\end{split}
		\end{equation}
		Applying Bernstein Inequality yields
		\[
			\mathbb{P}\left(\Bigl | \vec(\bu_1\bv_1^T)^T(\widehat\bSigma_{\tilde{\bX}\tilde{\bX}}- \bSigma_{\tilde{\bX}\tilde{\bX}})\vec(\bu_2\bv_2^T) \Bigr |> t\right) \le 2\exp\left(-c\min\Bigl (\frac{nt^2}{M^2\kappa_0^4}, \frac{nt}{M\kappa_0^2}\Bigr)\right).
		\]
		Finally, by the union bound over $(\bu_1, \bu_2, \bv_1, \bv_2)\in \cN^{d-1}\times \cN^{d-1}\times\cN^{d-1}\times \cN^{d-1}$, we have
\begin{equation}
\mathbb{P}\left(\Bigl | \sup_{\substack{\bu_1, \bu_2\in \cS^{d-1} \\ \bv_1, \bv_2 \in \cS^{d-1}} } \vec(\bu_1\bv_1^T)^T(\widehat\bSigma_{\tilde{\bX}\tilde{\bX}}- \bSigma_{\tilde{\bX}\tilde{\bX}})\vec(\bu_2\bv_2^T) \Bigr |> t\right) \le\exp\left(2d\log8-c\min\Bigl (\frac{nt^2}{M^2\kappa_0^4}, \frac{nt}{M\kappa_0^2}\Bigr)\right).
\end{equation}		
		
Take $t\asymp\sqrt{d/n}$, we derive the inequality (\ref{eq:6.33}).  By combining (\ref{eq:5.14}) and (\ref{eq:6.33}), we successfully prove (\ref{eq:6.30}).
		
\item RSC at $ \cL_n(\bTheta^*)$ over $\mathcal{C}(\mathcal{M}_r,\overline{\mathcal{M}}_r^{\perp},\bTheta^*)$

For all $\bDelta\in\mathcal{C}(\mathcal{M}_r,\overline{\mathcal{M}}_r^{\perp},\bTheta^*)=\left\{\bDelta\in\mathbb{R}^{d\times d}: \norm{\bDelta_{\overline{\mathcal{M}}_r^{\perp}}}_N\leq3\norm{\bDelta_{\overline{\mathcal{M}}_r}}_N+4\sum\limits_{j\geq r+1}\sigma_j(\bTheta^*)\right\}$ where $1\leq r \leq d$, we have

\begin{equation}\label{eq:5.19}
\norm{\bDelta}_N\leq\norm{\bDelta_{\overline{\mathcal{M}}_r}}_N+\norm{\bDelta_{\overline{\mathcal{M}}_r^{\perp}}}_N\leq 4\norm{\bDelta_{\overline{\mathcal{M}}_r}}_N+4\sum\limits_{j\geq r+1}\sigma_j(\bTheta^*)\leq4\sqrt{2r}\norm{\bDelta}_F+4\sum\limits_{j\geq r+1}\sigma_j(\bTheta^*)
\end{equation}

Let $\tilde{\kappa}=(1/8)\kappa$. As we did in the proof for Theorem \ref{thm:1}, we take $\tau=\lambda/\tilde{\kappa}$ and let $r=\#\{j\in\{1,2,...,d\}|\sigma_j(\bTheta^*)>\tau\}$. Then,
\begin{equation}\label{eq:418}
\sum_{j\geq r+1}\sigma_j(\bTheta^*)=\tau\cdot\sum_{j\geq r+1}\frac{\sigma_j(\bTheta^*)}{\tau}\leq\tau\cdot\sum_{j\geq r+1}{\frac{\sigma_j(\bTheta^*)}{\tau}}^q\leq\tau^{1-q}\rho=\lambda^{1-q}\tilde{\kappa}^{q-1}\rho
\end{equation}

On the other hand, $\rho>\sum_{j\leq r}\sigma(\bTheta^*)^q\geq r\tau^q$ so that $r\leq\rho\tau^{-q}=\rho\tilde{\kappa}^q\lambda^{-q}$. Plugging these results into (\ref{eq:5.19}), we have
\begin{equation}\label{eq:5.21}
\nnorm{\bDelta} \leq4\sqrt{2\rho}\lambda^{-q/2}\tilde{\kappa}^{q/2}\norm{\bDelta}_F+4\lambda^{1-q}\tilde{\kappa}^{q-1}\rho.
\end{equation}
Since $\lambda=2\nu\sqrt{d/n}$, there exist constants $c_5$ and $c_6$ such that as long as $\rho(d/n)^{(1-q) / 2}\leq c_4$, combining (\ref{eq:5.14}) and (\ref{eq:5.21}) we have
\begin{equation}\label{eq:6.41}
\text{vec}(\bDelta)^T\hat{\bH}(\bTheta^*)\text{vec}(\bDelta)\geq\tilde{\kappa}\norm{\bDelta}_F^2 - c_5 \rho \lambda^{2-q}.
\end{equation}
with high probability.

In the first two parts of this proof, we not only verify the RSC of $ \cL_n(\bTheta^*)$, but also provide the complete procedure of how to verify the RSC of the empirical loss given the RSC of the population loss. This is very important in Part 3 of this proof.

\item LRSC of $ \cL_n(\bTheta)$ around $\bTheta^*$

In the remaining proof, we verify the LRSC by showing that there exists a positive constant $\tilde{\kappa}'$ such that
\begin{equation}\label{eq:5.23}
\text{vec}(\hat{\bDelta})^T\hat{\bH}(\bTheta)\text{vec}(\hat{\bDelta})\geq\tilde{\kappa}'\norm{\hat{\bDelta}}_F^2 - c_6\rho \lambda^{2 - q}.
\end{equation}
holds for all $\widehat\bDelta\in\mathcal{C}(\mathcal{M}_r,\overline{\mathcal{M}}_r^{\perp},\bTheta^*)$ and $\bTheta$ such that $\fnorm{\bTheta-\bTheta^*}\leq c_7\sqrt{\rho}\lambda^{(1-q) / 2}$ for some positive constant $c_7$. Note that given ${\bTheta - \bTheta^*} \in \cC(\cM, \overline\cM^{\perp}, \bTheta^*)$, by \eqref{eq:5.21} we have $\nnorm{\bTheta - \bTheta^*} \le c_8 \rho \lambda^{1-q} =: \ell$ for some constant $c_8$.

Define functions $\hat{\bh}(\bTheta)=n^{-1}\sum\limits_{i=1}^nb''(\langle\bTheta, \bX_i\rangle)\cdot\mathds{1}_{\{\abs{\inn{\bTheta^*,\bX_i}}>\tau\norm{\bX_i}_{\text{op}}\}}\cdot\text{vec}(\bX_i)\text{vec}(\bX_i)^T$ and $\bh(\bTheta)=\mathbb{E}(\hat{\bh}(\bTheta))$ for constants $\tau$ and $\gamma$ to be determined. Recall that \[\hat{\bH}(\bTheta^*)=n^{-1}\sum\limits_{i=1}^nb''(\langle\bTheta^*, \bX_i\rangle)\text{vec}(\bX_i)\text{vec}(\bX_i)^T.\] The only difference between $\bh(\cdot)$ and $\bH(\cdot)$ is the indicator function so that $\widehat\bH(\cdot)\succeq\widehat\bh(\cdot)$.

We will finish the proof of LRSC in two steps. Firstly, we show that $\bh(\bTheta^*)$ is positive definite over the restricted cone. Then by following the procedure of showing (\ref{eq:6.41}), we can prove that $\hat{\bh}(\bTheta^*)$ is positive definite over the cone with high probability. Secondly, we bound the difference between $\text{vec}(\hat{\bDelta})^T\hat{\bh}(\bTheta)\text{vec}(\hat{\bDelta})$ and $\text{vec}(\hat{\bDelta})^T\hat{\bh}(\bTheta^*)\text{vec}(\hat{\bDelta})$ and show that $\hat{\bh}(\bTheta)$ is locally positive definite around $\bTheta^*$. This naturally lead to the LRSC of $ \cL_n(\bTheta)$ around $\bTheta^*$.

We establish the following lemma before proceeding.

\begin{lem}\label{lem:7}
When $\norm{\bTheta^*}_F\geq\alpha\sqrt{d}$ and $\{\vec{(\bX_i)}\}_{i=1}^n$ are sub-Gaussian, there exist a universal constant $\tau>0$ such that 
$\lambda_{\text{min}}(\bh(\bTheta^*))\geq\kappa_1$ where $\kappa_1$ is a positive constant.
\end{lem}


We select an appropriate $\tau$ to make $\bh(\bTheta^*)$ positive definite. Follow the same procedure in Part 1 and Part 2 of this proof, we derive that

\begin{equation}\label{eq:422}
\text{vec}(\hat{\bDelta})^T\cdot\hat{\bh}(\bTheta)\cdot\text{vec}(\hat{\bDelta})\geq\tilde{\kappa_1}\norm{\hat{\bDelta}}_F^2  - c_6\rho \lambda^{2 - q}.
\end{equation}
for a positive $\tilde{\kappa_1}$ with high probability.


Meanwhile,
\begin{equation}\label{eq:446}
\begin{split}
&\abs{\text{vec}(\hat{\bDelta})^T\cdot\hat{\bh}(\bTheta^*)\cdot\text{vec}(\hat{\bDelta})-\text{vec}(\hat{\bDelta})^T\cdot\hat{\bh}(\bTheta)\cdot\text{vec}(\hat{\bDelta})} \\
\leq &\text{vec}(\hat{\bDelta})^T\cdot\frac{1}{n}\sum\limits_{i=1}^n\abs{b''(\langle\bTheta^*, \bX_i\rangle)-b''(\langle\bTheta, \bX_i\rangle)}\cdot\mathds{1}_{\{\abs{\inn{\bTheta^*,\bX_i}}>\tau\norm{\bX_i}_{\text{op}}\}}\cdot\text{vec}(\bX_i)\text{vec}(\bX_i)^T\cdot\text{vec}(\hat{\bDelta}) \\
=& \text{vec}(\hat{\bDelta})^T\cdot\frac{1}{n}\sum\limits_{i=1}^n\abs{b'''(\langle\tilde{\bTheta}, \bX_i\rangle)\inn{\bTheta-\bTheta^*,\bX_i}}\cdot\mathds{1}_{\{\abs{\inn{\bTheta^*,\bX_i}}>\tau\norm{\bX_i}_{\text{op}}\}}\cdot\text{vec}(\bX_i)\text{vec}(\bX_i)^T\cdot\text{vec}(\hat{\bDelta})
\end{split}
\end{equation}

Here $\tilde{\bTheta}$ is a middle point between $\bTheta^*$ and $\bTheta$, thus it is also in the nuclear ball centered at $\bTheta^*$ with radius $\ell$. We know that $\abs{\inn{\tilde{\bTheta},\bX_i}}\geq\abs{\inn{\bTheta^*,\bX_i}}-\abs{\inn{\bTheta^*-\tilde{\bTheta},\bX_i}}\geq(\tau-\ell)\norm{\bX_i}_{\text{op}}$ when the indicator function equals to 1. If $\abs{\inn{\tilde{\bTheta},\bX_i}}>1$, according to Condition (C5),
\begin{equation}
\begin{split}
&\abs{b'''(\langle\tilde{\bTheta}, \bX_i\rangle)\inn{\bTheta-\bTheta^*,\bX_i}}\cdot\mathds{1}_{\{\abs{\inn{\bTheta^*,\bX_i}}>\tau\norm{\bX_i}_{\text{op}}\}} \\
\leq &\frac{\norm{\bX_i}_{\text{op}}\norm{\bTheta-\bTheta^*}_N}{\abs{\inn{\tilde{\bTheta},\bX_i}}}\leq\frac{\norm{\bX_i}_{\text{op}}\norm{\bTheta-\bTheta^*}_N}{(\tau-\ell)\norm{\bX_i}_{\text{op}}}\leq\frac{\ell}{\tau-\ell}
\end{split}
\end{equation}
Otherwise, $\norm{\bX_i}_{\text{op}}$ is bounded by $1/(\tau-\ell)$ and $\abs{b'''(\langle\tilde{\bTheta}, \bX_i\rangle)\inn{\bTheta-\bTheta^*,\bX_i}}\mathds{1}_{\{\abs{\inn{\bTheta^*,\bX_i}}>\tau\norm{\bX_i}_{\text{op}}\}}\leq C\cdot\frac{\ell}{\tau-\ell}$ where $C$ is the upper bound of $b'''(x)$ for $\abs{x}\leq1$. In summary,
\begin{equation}
(\ref{eq:446})\leq \text{vec}(\hat{\bDelta})^T\cdot\frac{C_1\ell}{n(\tau-\ell)}\sum\limits_{i=1}^n\text{vec}(\bX_i)\text{vec}(\bX_i)^T\cdot\text{vec}(\hat{\bDelta})
\end{equation}

where $C_1=\max(C,1)$. Denote $\hat{\bSigma}_{\bX\bX}=n^{-1}\sum\limits_{i=1}^n\vec{(\bX_i)}\vec{(\bX_i)}^T$ and $\bSigma_{\bX\bX}=\mathbb{E}\hat{\bSigma}_{\bX\bX}$. Suppose the eigenvalues of $\bSigma_{\bX\bX}$ is upper bounded by $K<\infty$, as a similar result to (\ref{eq:6.30}) and (\ref{eq:5.21}), as long as $\rho(d/n)^{1-q/2}\leq c_5$, we shall have
\begin{equation}\label{eq:425}
\begin{split}
 & \text{vec}(\hat{\bDelta})^T\cdot\frac{C_1\ell}{n(\tau-\ell)}\sum\limits_{i=1}^n\text{vec}(\bX_i)\text{vec}(\bX_i)^T\cdot\text{vec}(\hat{\bDelta}) \\
 \leq & \frac{C_1\ell}{(\tau-\ell)}\left(K\norm{\hat{\bDelta}}_F^2+C_0\sqrt{\frac{d}{n}}\norm{\hat{\bDelta}}_N^2\right) \\
\leq & \frac{2KC_1\ell}{\tau-\ell}\norm{\hat{\bDelta}}_F^2 \\
\end{split}
\end{equation}
with high probability. As long as the constant $\ell$ is sufficiently small such that $2KC_1\ell/(\tau-\ell)<\tilde{\kappa_1}/2$, $\text{vec}(\hat{\bDelta})^T\cdot\hat{\bh}(\bTheta)\cdot\text{vec}(\hat{\bDelta})\geq\tilde{\kappa}_2\fnorm{\hat{\bDelta}}^2$ holds with $\tilde{\kappa_2}=\tilde{\kappa_1}/2$. This delivers that $\widehat\bh(\bTheta)$ is locally positive definite around $\bTheta^*$ with hight probability. Recall that $\bH(\cdot)\succeq\bh(\cdot)$, we have verified that $\widehat\bH(\bTheta)$ is also locally positive definite around $\bTheta^*$. In summary, there exist some constant $\ell>0$ such that for any $\norm{\bTheta-\bTheta^*}_N\leq \ell$,
\begin{equation}
\text{vec}(\hat{\bDelta})^T\cdot\frac{1}{n}\sum\limits_{i=1}^nb''(\langle\bTheta, \bX_i\rangle)\text{vec}(\bX_i)\text{vec}(\bX_i)^T\cdot\text{vec}(\hat{\bDelta})\geq\tilde{\kappa}_2\norm{\hat{\bDelta}}_F^2 - c_6\rho\lambda^{2-q}.
\end{equation}
for all $\widehat\bDelta\in\mathcal{C}(\mathcal{M}_r,\overline{\mathcal{M}}_r^{\perp},\bTheta^*)$ with high probability. This finalized our proof of the LRSC of $ \cL_n(\bTheta)$ around $\bTheta^*$.

Below we provide the proof of Lemma \ref{lem:7}.

\underbar{\bf Proof for Lemma \ref{lem:7}}

We first show that for any $p_0\in(0,1)$, there exist constants $\tau$ and $\gamma$ such that $\mathbb{P}(\abs{\inn{\bTheta^*,\bX_i}}>\tau\norm{\bX_i}_{\text{op}})\geq p_0$.





We would show that $\mathbb{P}(\abs{\inn{\bTheta,\bX_i}}>c_1\sqrt{d})\geq (p_0+1)/2$ and $\mathbb{P}(\norm{\bX_i}_{\text{op}}\leq c_2\sqrt{d})\geq (p_0+1)/2$ for some positive constants $c_1$ and $c_2$. Then
\begin{equation}\label{eq:6.49}
\mathbb{P}(\abs{\inn{\bTheta,\bX_i}}>c_1/c_2\norm{\bX_i}_{\text{op}})\geq (p_0+1)/2+(p_0+1)/2-1=p_0
\end{equation}

On one hand, $\inn{\bTheta,\bX_i}$ is a sub-Gaussian variable since it is a linear transformation of a sub-Gaussian vector. Its mean is 0 and its sub-Gaussian norm is bounded by $\kappa_0\norm{\bTheta}_F$. Since $\norm{\bTheta}_F\geq \alpha\sqrt{d}$, take $c_1$ to be sufficiently small, we have
\begin{equation}
\mathbb{P}(\abs{\inn{\bTheta,\bX_i}}>c_1\sqrt{d})\geq\mathbb{P}(\abs{x}>c_1/\alpha)\geq \frac{p_0+1}{2}
\end{equation}
where $x$ is a sub-Gaussian variable and $\norm{x}_{\psi_2}\leq\kappa_0$.

On the other hand,
\begin{equation}
\begin{split}
\norm{\bX_i}_{\text{op}}=&\max\limits_{\bu\in\cS^d,\bv\in\cS^d}\abs{\bu^T\bX_i\bv}=\max\limits_{\bu\in\cS^d,\bv\in\cS^d}\abs{\text{tr}(\bu^T\bX_i\bv)} \\
=&\max\limits_{\bu\in\cS^d,\bv\in\cS^d}\abs{\text{tr}(\bv\bu^T\bX_i)}=\max\limits_{\bu\in\cS^d,\bv\in\cS^d}\abs{\inn{\bu\bv^T,\bX_i}}.
\end{split}
\end{equation}

Recall the covering argument in the proof of Lemma \ref{lem:1}. Denote $\mathcal{N}^d$ as a $1/4$-net on $\cS^d$, then
\begin{equation}
\max\limits_{\bu\in\cS^d,\bv\in\cS^d}\abs{\inn{\bu\bv^T,\bX_i}}\leq \frac{16}{7}\max\limits_{\bu\in\mathcal{N}^d,\bv\in\mathcal{N}^d}\abs{\inn{\bu\bv^T,\bX_i}}
\end{equation}

For any $\bu_1\in\mathcal{N}^d$, $\bv_1\in\cN^d$, given $\norm{\bX_i}_{\psi_2}\leq\kappa_0$, we have $\norm{\inn{\bu_1\bv_1^T,\bX_i}}_{\psi_1}\leq \kappa_0$. According to Bernstein-type inequality in Vershynin (2010), it follows that for sufficiently small $t$ and some positive constant $C$,
\begin{equation}
\mathbb{P}(\abs{\inn{\bu_1\bv_1^T,\bX_i}}>t)\leq2\exp\left(-\frac{Ct^2}{\kappa_0^2}\right)
\end{equation}

Therefore, the overall union bound follows:
\begin{equation}
\mathbb{P}(\max\limits_{\bu\in\cS^d,\bv\in\cS^d}\abs{\inn{\bu\bv^T,\bX_i}}>t)\leq 2\exp\left(2d\log{4}-\frac{Ct^2}{\kappa_0^2}\right)
\end{equation}

Let $t=c_2\sqrt{d}$ for some positive constant $c_2>\sqrt{4\log{4}\kappa_0^2/C}$, the above probability decays. This means that with high probability (which is greater than $(p_0+1)/2$) $\norm{\bX_i}_{\text{op}}$ is less than $c_2\sqrt{d}$. This finalize our proof of (\ref{eq:6.49}).

Now we look at $\bh(\bTheta)=n^{-1}\mathbb{E}\left[\sum\limits_{i=1}^nb''(\langle\bTheta, \bX_i\rangle)\cdot\mathds{1}_{\{\abs{\inn{\bTheta^*,\bX_i}}>\tau\norm{\bX_i}_{\text{op}}\}}\cdot\text{vec}(\bX_i)\text{vec}(\bX_i)^T\right]$. Denote $\{\abs{\inn{\bTheta^*,\bX_i}}>\tau\norm{\bX_i}_{\text{op}}\}$ as an event $A_i$ with probability sufficiently close to 1. For any $\bv\in\mathbb{R}^{d^2}$,
\begin{equation}
\begin{split}
n\bv^T\bh(\bTheta^*)\bv=&\mathbb{E}\left[\sum\limits_{i=1}^nb''(\langle\bTheta^*, \bX_i\rangle)(\text{vec}(\bX_i)^T\bv)^2\right]-\mathbb{E}\left[\sum\limits_{i=1}^nb''(\langle\bTheta^*, \bX_i\rangle)\cdot\mathds{1}_{A_i^c}\cdot(\text{vec}(\bX_i)^T\bv)^2\right] \\
\geq &n\kappa\norm{\bv}_2^2- \sqrt{\mathbb{E}\left[\sum\limits_{i=1}^nb''(\langle\bTheta^*, \bX_i\rangle)^2\left(\text{vec}(\bX_i)^T\bv\right)^4\right]}\cdot\sqrt{\mathbb{E}\sum\limits_{i=1}^n\mathds{1}_{A_i^c}}\\
\geq &n\kappa\norm{\bv}_2^2-nMK\sqrt{1-p_0}\norm{\bv}_2^2
\end{split}
\end{equation}
Here, $M$ is an global upper bound of $b''(\cdot)$ and $K$ is the largest eigenvalue of the fourth moment of $\bX_i$. Since $\bX_i$ is sub-Gaussian, the fourth moment is bounded. We let $1-p_0$ be sufficiently small so that $nMK\sqrt{1-p_0}\leq\kappa/2$, then we proved that $\lambda_{\text{min}}(\bh(\bTheta^*))\geq\kappa/2>0$ and thus $\bh(\bTheta^*)$ is positive definite.

\end{enumerate}

\subsection{\bf Proof of Lemma \ref{lem:3}}

\[
	\begin{aligned}
		\frac{1}{N}\sum\limits_{i=1}^N (b'(\inn{\bX_i, \bTheta^*}) - Y_i)\bX_i & = \frac{1}{n} \sum\limits_{i=1}^{n} \frac{1}{d}\sum\limits_{j=1}^d (b'({\btheta_j^*}^T\bx_i) - y_{ij})\bx_i\be_j^T = \frac{1}{d} \cdot \frac{1}{n}\sum\limits_{i=1}^n  \bx_i\bz_i^T,
	\end{aligned}
\]
where $\bz_i$ satisfies that $z_{ij} = b'({\btheta^*_j}^T\bx_i) - y_{ij}$. Note that given $\bx_i$, $\|z_{ij}\|_{\psi_2} \le \phi M$. To see why, let $\eta_{ij} = \bx_i^T \btheta_j^*$. We have
\[	
			\begin{aligned}
			\E \exp(tz_{ij} \given \bx_i) & = \int_{y\in \cY}c(y)\exp\bigl(\frac{\eta_{ij} y- b(\eta_{ij})}{\phi}\bigr)\exp(t(y - b'(\eta_{ij})))dy \\
			& =\int_{y \in \cY} c(y)\exp\bigl(\frac{(\eta_{ij}+\phi t)y- b(\eta_{ij}+\phi t)+ b(\eta_{ij}+\phi t)- b(\eta_{ij})- \phi t b'(\eta_{ij})}{\phi}\bigr) dy \\
			& = \exp\bigl( \frac{b(\eta_{ij}+ \phi t)- b(\eta_{ij})- \phi t b'(\eta_{ij})}{\phi}\bigr) \le \exp\bigl(\frac{\phi M t^2}{2} \bigr).
			\end{aligned}
\]
Besides, $y_{ij}\independent y_{ik}$ for $j \neq k$ given $\bx_i$. Therefore, $\|\bz_i\|_{\psi_2} \le \phi M $. Since $ \E \bz_i\bx_i^T = \bzero$, by the standard covering argument used in the proof of \ref{lem:1}, for any $\nu>0$, there exists $\gamma > 0$ such that when $n > \gamma d$, it holds for some constant $c>0$,
\[
 	\PP\bigl(\opnorm{\frac{1}{n}\sum \limits_{i=1}^n \bx_i\bz_i^T} \ge \nu\sqrt{\frac{\phi M \kappa_0 d}{n}}\bigr) \le 2\exp(-cd).
\]

\subsection {\bf Proof of Lemma \ref{lem:4}}
\beq	
			\label{eq:6.38}
			\begin{aligned}
				\vec(\widehat\bDelta)^T \widehat \bH(\bTheta^*)\vec(\widehat\bDelta) & =\frac{1}{N}\sum\limits_{i=1}^N b''(\inn{\bX_i, \bTheta^*})\inn{\widehat\bDelta, \bX_i}^2= \frac{1}{N}\sum\limits_{i=1}^n\sum\limits_{j=1}^{d} b''(\bx_i^T\btheta^*_j)\inn{\widehat\bDelta, \bx_j\be_i^T}^2 \\
				& = \frac{1}{N} \sum\limits_{i=1}^n \sum\limits_{j=1}^{d} b''(\bx_i^T\btheta^*_j)\tr (\bx_i^T\widehat\bDelta \be_j)^2 = \frac{1}{N} \sum\limits_{i=1}^n \sum\limits_{j=1}^{d} b''(\bx_i^T\btheta^*_j) (\bx_i^T\widehat\bDelta_j)^2.
			\end{aligned}
\eeq

Note that for any $1 \le j \le d$, $\|\sqrt{b''(\bx_i^T \btheta_j)\bx_i}\|_{\psi_2}\le \sqrt{M}\kappa_0$. By Theorem 5.39 in \cite{Ver10}, there exists some $\gamma > 0$ such that if $n > \gamma d$, we have for some universal constant $c >0$,
\beq
	\label{eq:6.59}
	\mathbb{P}\Bigl( \opnorm{\frac{1}{n}\sum\limits_{i=1}^n b''(\bx_i^T\btheta^*_j) \bx_i\bx_i^T - \E  (b''(\bx_i^T\btheta^*_j)\bx_i\bx_i^T)}  \ge \kappa_0\sqrt{\frac{Md}{n}} \Bigr) \le 2 \exp(-cd).
\eeq
By the union bound, it holds that
\[
	\PP\Bigl( \max_{1 \le j \le d} \opnorm{\frac{1}{n}\sum\limits_{i=1}^n b''(\bx_i^T\btheta^*_j) \bx_i\bx_i^T - \bH(\bTheta^*)} \ge \kappa_0\sqrt{\frac{d}{n}}\Bigr) \le 2d \exp(-cd).
\]
In addition, for any $\bTheta \in \RR^{d \times d}$ such that $\fnorm{\bTheta - \bTheta^*} \le r$, $\ltwonorm{\btheta_j - \btheta_j^*} \le r$ holds for all $1 \le j \le d$. Given that $\|\bx_i\|_{\psi_2} \le \kappa_0$,
\[
	\PP(\max_{1 \le i \le n, 1 \le j \le d} |\bx_i^T (\btheta_j - \btheta^*_j)| \ge t ) \le 2nd\exp\Bigl( -\frac{t^2}{2\kappa^2_0 r^2} \Bigr).
\]
Substituting $t = \kappa_0r\sqrt{\delta \log(nd)}$ into the inequality above, we have
\[
	\PP(\max_{1 \le i \le n, 1 \le j \le d} |\bx_i^T (\btheta_j - \btheta^*_j)| \ge \kappa_0r \sqrt{\delta \log(nd)} ) \le 2(nd)^{1 - \frac{\delta}{2}}.
\]
Denote the above event by $\cE_1$. Therefore, under $\cE_1^c$,
\beq
	\label{eq:6.60}
	\begin{aligned}
	\opnorm{\frac{1}{n}\sum\limits_{i=1}^n (b''(\bx_i^T\btheta_j) -  b''(\bx_i^T\btheta^*_j))\bx_i\bx_i^T} & \le L \opnorm{\frac{1}{n} \sum\limits_{i=1}^n (\bx_i^T(\btheta_j - \btheta_j^*)) \bx_i\bx_i^T} \\
	& \le L \kappa_0r\sqrt{\delta \log(nd)} \cdot \opnorm{\frac{1}{n} \sum\limits_{i=1}^n \bx_i\bx_i^T}.
	\end{aligned}
\eeq
Again by Theorem 5.39 in \cite{Ver10}, when $n / d$ is sufficiently large,
\[
	\mathbb{P}\Bigl( \opnorm{\frac{1}{n}\sum\limits_{i=1}^n \bx_i\bx_i^T - \bSigma_{\bx\bx}}  \ge \kappa_0\sqrt{\frac{td}{n}} \Bigr) \le 2 \exp(-cd).
\]
Therefore, when $n / d$ is sufficiently large, $\opnorm{n^{-1} \sum\limits_{i=1}^n \bx_i\bx_i^T} \le 2\kappa_0$. Denote this event by $\cE_2$. Combining this with \eqref{eq:6.59} and \eqref{eq:6.60}, we have under $\cE^c_1 \cap \cE^c_2$,
\[
	\opnorm{\frac{1}{n}\sum\limits_{i=1}^n (b''(\bx_i^T\btheta_j) -  b''(\bx_i^T\btheta^*_j))\bx_i\bx_i^T} \le  2L\kappa^2_0 r \sqrt{\delta \log(nd)}.
\]
Finally, for sufficiently large $n / d$, it holds with probability at least $1 - 2(nd)^{1 - \frac{\delta}{2}}$ for all $\btheta$ such that $\fnorm{\bTheta - \bTheta^*} \le r$,
\[	
	\lambda_{\min} \Bigl( \frac{1}{n}\sum\limits_{i=1}^n b''(\bx_i^T\btheta_j)\bx_i\bx_i^T \Bigr) \ge \kappa_{\ell} - 2L\kappa^2_0 r \sqrt{\delta \log(nd)}.
\]
By a union bound across $j = 1, \ldots, d$, we can deduce that for any $\delta >4$, it holds with probability at least $1 - 2(nd)^{2 - \frac{\delta}{2}}$ that for all $\bDelta\in\mathbb{R}^{d\times d}$ and all $\bTheta \in \cN$,
\[
\text{vec}(\bDelta)^T \widehat{\bH}(\bTheta) \text{vec}(\bDelta)\geq \frac{1}{d} (\kappa_{\ell} -  2L\kappa^2_0 r \sqrt{\delta \log(nd)}) \fnorm{\bDelta}^2.
\]
Since $r \asymp {\sqrt{\rho} \lambda^{1- q/ 2}}$, as long as $\rho(d / n)^{1 - q/ 2}\log(nd)$ is sufficiently small, LRSC$(\cC, \cN, (1 / 2)\kappa_{\ell}, 0)$ holds.

\subsection{\bf Proof for Lemma \ref{lem:5}}

Here, we take advantage of the singleton design of $X$ and apply the Matrix Bernstein inequality (Theorem 6.1.1 in Tropp(2015)) to bound the operator norm of the gradient of the loss function.

Denote $\bZ_i=\left[\exp{(\inn{\bTheta^*,\bX_i})}/(1+\exp{(\inn{\bTheta^*,\bX_i})})-Y_i\right]\cdot\bX_i\in\mathbb{R}^{d\times d}$. $\forall \bu\in\cS^d, \bv\in\cS^d$,
\[\bu^T\bZ_i\bv\leq\ \left|\frac{e^{\inn{\bTheta^*,X_i}}}{e^{\inn{\bTheta^*,X_i}}+1}-Y_i\right|\cdot d\leq d.\]

Thus $\norm{\bZ_i}_{\text{op}}\leq d$. Meanwhile,

\begin{equation}
\begin{split}
\norm{\mathbb{E}\bZ_i\bZ_i^T}_{\text{op}}=&\norm{\mathbb{E}\left[\left(\frac{e^{\inn{\bTheta^*,\bX_i}}}{e^{\inn{\bTheta^*,\bX_i}}+1}-Y_i\right)^2\bX_i\bX_i^T\right]}_{\text{op}}\leq\norm{\mathbb{E}\left[\bX_i\bX_i^T\right]}_{\text{op}} \\
=&d^2\cdot\norm{\mathbb{E}\left[\be_{a(i)}\be_{a(i)}^T\right]}_{\text{op}}=d^2\cdot\frac{1}{d}=d \\
\end{split}
\end{equation}

Similarly, we have $\norm{\mathbb{E}\bZ_i^T\bZ_i}_{\text{op}}\leq d$. Therefore, $\text{max}\left\{\norm{\mathbb{E}\bZ_i\bZ_i^T}_{\text{op}},\norm{\mathbb{E}\bZ_i^T\bZ_i}_{\text{op}}\right\}\leq d$.

According to Matrix Bernstein inequality,

\begin{equation}
P\left(\norm{\frac{1}{n}\sum\limits_{i=1}^n \bZ_i}_{\text{op}}\geq t\right)\leq2d\cdot\exp{(\frac{-nt^2/2}{d+dt/3})}
\end{equation}

Let $t=\nu\sqrt{\delta d\log{d}/n}$, then

\begin{equation}
\begin{split}
P\left(\norm{\frac{1}{n}\sum\limits_{i=1}^n \bZ_i}_{\text{op}}\geq \nu\sqrt{\frac{\delta d\log{d}}{n}}\right)\leq &2d\cdot\exp{(\frac{-\nu^2\delta d\log{d}}{2d+2\nu\sqrt{\frac{d^2\delta d\log{d}}{n}}/3})} \\
=&2d^{1-\frac{\nu^2\delta}{2+2\nu\sqrt{d\cdot\delta\cdot\log{d}}/3\sqrt{n}}} \\
\leq & 2d^{1-\delta}
\end{split}
\end{equation}

for some constant $\nu$ as long as $d\log{d}/n\leq\gamma$ for some constant $\gamma$.

\subsection{Proof for Lemma \ref{lem:6}}

We aim to show that the loss function has LRSC property in a $L_{\infty}$-ball centered at $\bTheta^*$ with radius $2R/d$.

For all $\tilde{\bTheta}\in\mathbb{R}^{d\times d}$ satisfying $\norm{\tilde{\bTheta}-\bTheta^*}_{\infty}\leq 2R/d$, let us denote $f(\bTheta)=\exp{(\inn{\bTheta^*,\bX_i})}/(1+\exp{(\inn{\bTheta^*,\bX_i})})^2$. Then
\begin{equation}
\begin{split}
&\vec{({\bDelta})}^T[\hat{\bH}(\tilde{\bTheta})-\hat{\bH}(\bTheta^*)]\vec{({\bDelta})} \\
=&\vec{({\bDelta})}^T\cdot\frac{1}{n}\sum\limits_{i=1}^n \left[f\left(\inn{\tilde{\bTheta},\bX_i}\right)-f\left(\inn{\bTheta^*,\bX_i}\right)\right]\vec{(\bX_i)}\vec{(\bX_i)}^T\cdot\vec{({\bDelta})} \\
\leq &\vec{({\bDelta})}^T\cdot\frac{1}{n}\sum\limits_{i=1}^n f'\left(\inn{\bar{\bTheta}_i,\bX_i}\right)\inn{\tilde{\bTheta}-\bTheta^*,\bX_i}
\vec{(\bX_i)}\vec{(\bX_i)}^T\cdot\vec{({\bDelta})}
\end{split}
\end{equation}
Here $\bar{\bTheta}_i$ is a middle point between $\tilde{\bTheta}$ and $\bTheta^*$.
Due to the singleton design of $\bX_i$, $\inn{\tilde{\bTheta}-\bTheta^*,\bX_i}\leq d\cdot\norm{\tilde{\bTheta}-\bTheta^*}_{\infty}\leq 2R$. Given that the derivative of $f(\cdot)$ is bounded by 0.1, we have

\begin{equation}\label{eq:441}
\begin{split}
\vec{({\bDelta})}^T[\hat{\bH}(\tilde{\bTheta})-\hat{\bH}(\bTheta^*)]\vec{({\bDelta})}\leq &\frac{R}{5}\cdot\vec{({\bDelta})}^T\cdot\frac{1}{n}\sum\limits_{i=1}^n \vec{(\bX_i)}\vec{(\bX_i)}^T\cdot\vec{({\bDelta})} \\
=: &\frac{R}{5n}\norm{\tilde{\mathfrak{X}}_n({\bDelta})}_2^2
\end{split}
\end{equation}

It is proved in the proof of Theorem 1 in \cite{NWa12} that as long as $n>c_6d\log{d}$,

\begin{equation}
\abs{\frac{\norm{\tilde{\mathfrak{X}}_n(\bDelta)}_2}{\sqrt{n}}-\norm{\bDelta}_F}\geq\frac{7}{8}\norm{\bDelta}_F+\frac{16d\norm{\bDelta}_{\infty}}{\sqrt{n}}
\end{equation}
for all $\bDelta\in\cC'(c_0)$ with probability at most $c_7\exp{(-c_8d\log{d})}$. Therefore, since ${\bDelta}\in\cC'(c_0)$ and $128d\norm{{\bDelta}}_{\infty}/\sqrt{n}\norm{{\bDelta}}_F\leq1/2$, we shall have

\begin{equation}\label{eq:443}
\frac{\norm{\tilde{\mathfrak{X}}_n(\bDelta)}_2}{\sqrt{n}}\leq \frac{15}{8}\norm{{\bDelta}}_F+\frac{16d\norm{{\bDelta}}_{\infty}}{\sqrt{n}}\leq \left(\frac{15}{8}+\frac{1}{16}\right)\norm{{\bDelta}}_F\leq2\norm{{\bDelta}}_F
\end{equation}
with probability greater than $1-c_7\exp{(-c_8d\log{d})}$. When (\ref{eq:443}) holds, plug it into (\ref{eq:441}), we shall have

\begin{equation}\label{eq:620}
\vec{({\bDelta})}^T[\hat{\bH}(\tilde{\bTheta})-\hat{\bH}(\bTheta^*)]\vec{({\bDelta})}\leq\frac{R}{5}\cdot 4\norm{{\bDelta}}_F^2\leq\frac{\norm{{\bDelta}}_F^2}{512(e^R+e^{-R}+2)}
\end{equation}
for sufficiently small $R>0$. The following inequality thus holds for all $\tilde{\bTheta}$ satisfying $\norm{\tilde{\bTheta}-\bTheta^*}_{\infty}\leq 2R/d$:

\begin{equation}
\vec{({\bDelta})}^T\hat{\bH}(\tilde{\bTheta})\vec{({\bDelta})}\geq\frac{\norm{{
\bDelta}}_F^2}{512(e^R+e^{-R}+2)}
\end{equation}

\subsection{\bf Proof for Theorem \ref{thm:4}}

In this proof, we define an operator $\tilde{\mathfrak{X}}_n:\mathbb{R}^{d\times d}\rightarrow\mathbb{R}^n$ such that $[\tilde{\mathfrak{X}}_n(\bGamma)]_i=\inn{\bGamma,\bX_i}$ for all $\bGamma\in\mathbb{R}^{d\times d}$.

Denote $\hat{\bDelta}=\hat{\bTheta}-\bTheta^*$. If $\hat{\bDelta}\notin\cC'(c_0)$, according to Case 1 in the proof for Theorem 2 in \cite{NWa12}, we shall have

\begin{equation}
\norm{\hat{\bDelta}}_F^2\leq 2c_0R\sqrt{\frac{d\log{d}}{n}}\cdot\left\{8\sqrt{r}\norm{\hat{\bDelta}}_F+4\sum\limits_{j=r+1}^{d}\sigma_j(\bTheta^*)\right\}
\end{equation}
for any $1\leq r\leq d$. Following the same strategy we used in the proof for Theorem \ref{thm:1}, we will have \[\norm{\hat{\bDelta}}_F\leq C_1\sqrt{\rho}\left(2C_1R\sqrt{\frac{d\log{d}}{n}}\right)^{1-q/2}\]
for some constant $C_1$.

If $\hat{\bDelta}\in\cC'(c_0)$, when (\ref{eq:26}) in Lemma \ref{lem:1} holds, on one hand, if $128d\norm{\hat{\bDelta}}_{\infty}/\sqrt{n}\norm{\hat{\bDelta}}_F>1/2$, we have

\begin{equation}
\norm{\hat{\bDelta}}_F\leq\frac{256d\norm{\hat{\bDelta}}_{\infty}}{ \sqrt{n}}\leq\frac{512R}{\sqrt{n}}
\end{equation}

As what we did in the proof for Theorem \ref{thm:1}, we take $\tau=\left(R^2/\rho n\right)^{\frac{1}{2-q}}$ and we have

\begin{equation}
\norm{\hat{\bDelta}}_N\leq C_2\left(\rho\left(\frac{R^2}{n}\right)^{1-q}\right)^{\frac{1}{2-q}}
\end{equation}
for some constant $C_2$.

On the other hand, if $128d\norm{\hat{\bDelta}}_{\infty}/\sqrt{n}\norm{\hat{\bDelta}}_F\leq1/2$, we have

\begin{equation}\label{eq:439}
\frac{\norm{\mathfrak{X}_n(\hat{\bDelta})}_2}{\sqrt{n}}\geq\frac{\norm{\hat{\bDelta}}_F}{16(e^{R/2}+e^{-R/2})}\quad\text{i.e., }\quad\frac{\norm{\mathfrak{X}_n(\hat{\bDelta})}_2^2}{n}\geq\frac{\norm{\hat{\bDelta}}_F^2}{256(e^R+e^{-R}+2)}
\end{equation}

Thus by Lemma \ref{lem:1} and \ref{lem:2} it naturally holds that
\[\norm{\hat{\bTheta}-\bTheta}_F^2\leq C_3\rho\left(\sqrt{\frac{d\log{d}}{n}}\right)^{2-q},\quad\norm{\hat{\bTheta}-\bTheta}_N\leq C_4\rho\left(\sqrt{\frac{d\log{d}}{n}}\right)^{1-q}.\]

In summary, as long as $n / (d\log{d})$ is sufficiently large, we shall have
\begin{equation}
\begin{split}
&\norm{\hat{\bTheta}-\bTheta^*}_F^2\leq C_5\max\left\{\rho\left(\sqrt{\frac{d\log{d}}{n}}\right)^{2-q},\frac{R^2}{n}\right\},  \\
&\norm{\hat{\bTheta}-\bTheta^*}_N\leq C_6\max\left\{\rho\left(\sqrt{\frac{d\log{d}}{n}}\right)^{1-q},\left(\rho\left(\frac{R^2}{n}\right)^{1-q}\right)^{\frac{1}{2-q}}\right\} \\
\end{split}
\end{equation}
with probability greater than $1-C_7\exp{(-c_1d\log{d})}-2d^{1-\delta}$, where $\{C_i\}_{i=5}^7$ and $c_1$ are constants.

%
%
%
%
%
%
%

\end{document}